\documentclass[twocolumn,
amsmath,amssymb,aps,pra,floatfix]{revtex4-1}

\usepackage[subpreambles=true]{standalone}
\usepackage{import}

\usepackage{graphicx}
\usepackage{dcolumn}
\usepackage{bm}

\usepackage{comment}

\usepackage[colorlinks,urlcolor=blue,citecolor=blue,linkcolor=blue]{hyperref}
\usepackage[all]{hypcap}

\newcommand{\vect}[1]{{\mathbf #1}}

\renewcommand{\k}{{\bf k}}
\newcommand{\ekb}{\epsilon_{\k}^B}
\newcommand{\eqb}{\epsilon_{\q}^B}
\newcommand{\p}{{\bf p}}

\newcommand{\q}{{\bf q}}
\renewcommand{\l}{{\bf l}}

\newcommand{\0}{{\bf 0}}

\newcommand{\ab}{a_{B}}

\newcommand{\ket}[1]{\left|{#1}\right>}

\newcommand{\eb}{E_{\rm B}}
\newcommand{\eq}{\epsilon_{\q}}
\newcommand{\ek}{\epsilon_{\k}}

\newcommand{\nn}{\nonumber}
\newcommand{\beq}{\begin{equation}}
\newcommand{\eeq}{\end{equation}}

\newcommand{\nx}{n_{\rm ex}}

\newcommand{\sch}{Schr{\"o}dinger }

\newcommand{\cre}[2]{\hat{#1}_{#2}^{\dagger}}
\newcommand{\ann}[2]{\hat{#1}_{#2}}
\newcommand{\num}[2]{\cre{#1}{#2} \ann{#1}{#2}}
\newcommand{\abs}[1]{\left| #1 \right|}

\def\lba{\left(}    \def\rba{\right)}

\def \del{\partial}       

\newcommand{\sout}[1]{}

\begin{document}

\title{Fate of the Bose polaron at finite temperature}

\author{Bernard Field}
\affiliation{School of Physics and Astronomy, Monash University, Victoria 3800, Australia}
\affiliation{ARC Centre of Excellence in Future Low-Energy Electronics Technologies, Monash University, Victoria 3800, Australia}

\author{Jesper Levinsen}
\affiliation{School of Physics and Astronomy, Monash University, Victoria 3800, Australia}
\affiliation{ARC Centre of Excellence in Future Low-Energy Electronics Technologies, Monash University, Victoria 3800, Australia}

\author{Meera M.\ Parish}
\affiliation{School of Physics and Astronomy, Monash University, Victoria 3800, Australia}
\affiliation{ARC Centre of Excellence in Future Low-Energy Electronics Technologies, Monash University, Victoria 3800, Australia}

\date{\today}

\begin{abstract} 
We consider an impurity immersed in a Bose-Einstein condensate with tunable boson-impurity interactions. Such a Bose polaron has recently been predicted to exhibit an intriguing energy spectrum at finite temperature, where the ground-state quasiparticle evenly splits into two branches as the temperature is increased from zero [Guenther et al., Phys.~Rev.~Lett.~\textbf{120}, 050405 (2018)].  To investigate this theoretical prediction, we employ a recently developed variational approach that systematically includes multi-body correlations between the impurity and the finite-temperature medium, thus allowing us to go beyond previous finite-temperature methods.  Crucially, we find that the number of quasiparticle branches is simply set by the number of hole excitations of the thermal cloud, such that including up to one hole yields one splitting, two holes yields two splittings, and so on. Moreover, this effect is independent of the impurity mass.  We thus expect that the exact ground-state quasiparticle will evolve into a single broad peak for temperatures $T>0$, with a broadening that scales as $T^{3/4}$ at low temperatures and sufficiently weak boson-boson interactions.  In the zero-temperature limit, we show that our calculated ground-state polaron energy is in excellent agreement with recent quantum Monte Carlo results and with experiments.  
\end{abstract}


\maketitle

\section{Introduction}

The scenario of an impurity coupled to a quantum medium is ubiquitous in physics, having relevance to systems ranging from field-effect transistors~\cite{Hulea2006} to protons in neutron stars~\cite{PhysRevC.47.1077}.  Most recently, it has been realized in trapped cold atomic gases, which provide a particularly clean and flexible system in which to probe the behavior of quantum impurities~\cite{Massignan_Zaccanti_Bruun}.  Here, the impurity-medium interactions can be varied from weak to strong coupling via Feshbach resonances~\cite{Chin2010}, allowing one to investigate the manner in which the impurity becomes ``dressed'' by excitations of the medium.  Cold-atom experiments have already greatly improved our understanding of the impurity in a degenerate Fermi gas --- otherwise known as the Fermi polaron~\cite{Schirotzek2009,Nascimbene2009,Kohstall2012,Koschorreck2012,Cetina2015doi,Ong2015,Cetina2016,Scazza2017,Yan2019prl,Oppong2019}.  However, attention is now focusing more on the \textit{Bose polaron}, corresponding to the case of a mobile impurity in a Bose-Einstein condensate (BEC).

In the limit of weak boson-impurity interactions, the Bose polaron maps onto the Fr\"{o}hlich model~\cite{Tempere2009,Rentrop2016} and thus realizes the original concept of a polaron, where an electron couples to phonons in an ionic lattice~\cite{Pekar1946,Landau1948,Frohlich1954}.  However, the regime of strong boson-impurity coupling is far less understood and has only recently been accessed in cold-atom experiments, where long-lived quasiparticles were recently observed~\cite{Jorgensen2016,Hu2016,Yan2019}.  Moreover, unlike the Fermi polaron, the medium for the Bose polaron exhibits a phase transition from a BEC to a thermal Bose gas at finite temperature, which can dramatically change the character of the impurity quasiparticle~\cite{Levinsen2017,Pastukhov2018,Guenther2018}.

The Bose polaron problem at strong coupling has been tackled with a variety of theoretical techniques, including quantum Monte Carlo (QMC)~\cite{Ardila2015,Ardila2016}, field-theoretical diagrammatic approaches~\cite{Rath2013}, renormalization group theory~\cite{Grusdt2017}, and variational wave functions involving highly correlated~\cite{Li2014,Levinsen2015,Yoshida2018} and coherent-state~\cite{Shchadilova2016,Drescher2018,Loon2018} ansatzes.  However, the majority of the theoretical work so far has been restricted to zero temperature, where the impurity-medium system is in a pure state and is thus easier to treat.

For the finite-temperature Bose polaron, most of the theoretical investigations have been perturbative in nature, being only valid in the limit of weak boson-impurity interactions~\cite{Levinsen2017,Pastukhov2018,Lausch2018,Lampo2018} or at high temperatures well above the BEC critical temperature~\cite{Sun2017}.  While a non-perturbative functional determinant approach has been used for Rydberg atoms in a Bose gas~\cite{Schmidt2016,Camargo2018}, the Rydberg polaron differs significantly from the canonical Bose polaron where the impurity is point-like.  Recently, Guenther et al.~\cite{Guenther2018} developed a finite-temperature diagrammatic scheme that they used to compute the polaron energy spectrum at all coupling strengths and temperatures.  In the strong-coupling regime, this ``extended'' ladder approximation predicts that the ground-state polaron (the attractive polaron) splits into two quasiparticles as the temperature is increased from zero~\cite{Guenther2018}.  This surprising prediction has important consequences for the nature of the Bose polaron at finite temperature, but an open question is whether or not the observed quasiparticle splitting persists at higher orders in the approximation, beyond the ladder diagrams.

In this paper, we investigate this problem using a finite-temperature variational method, which allows us to successively include medium excitations in the description of the Bose polaron. Such an approach was recently formulated for the Fermi polaron at finite temperature~\cite{Liu2019}, where it was shown to successfully model recent cold-atom experiments on impurity dynamics~\cite{Cetina2016}.  Unlike other many-body methods, it exactly captures few-body correlations such as three-body Efimov physics~\cite{Levinsen2015,Yoshida2018}, which is an important feature of any three-dimensional Bose system with short-range interactions. In particular, by carefully identifying the Efimov length scale, we show that our approach accurately predicts the zero-temperature ground-state energy measured in experiment~\cite{Jorgensen2016} and calculated using quantum Monte Carlo~\cite{Pena-Ardila2019}. Our observation thus provides further evidence that the ground-state energy of the Bose polaron in the strongly interacting regime is a universal function of the Efimov length scale, as argued in Ref.~\cite{Yoshida2018}.

At finite temperature, we investigate the full spectral function of the Bose polaron within our variational approach~\cite{Liu2019}. Including all two-body correlations involving the impurity and a bosonic excitation, we recover the results of Ref.~\cite{Guenther2018}; namely, we observe the splitting of the attractive polaron into two branches at temperature $0<T<T_c$, where $T_c$ is the critical temperature for Bose-Einstein condensation. However, our variational approach naturally allows us to go beyond the usual ladder approximation, and including all possible three-body correlations consisting of the impurity and two bosonic excitations, we find that the attractive polaron splits into three branches.

We argue that the attractive polaron splitting arises due to the two-fluid nature of the finite-temperature Bose gas, where the impurity can scatter particles between the two fluids. Using a diagrammatic analysis, we show that in general the number of splittings will equal the number of hole excitations included in the variational polaron ansatz. As a consequence, we show that the pole condition of the attractive polaron at finite temperature involves an infinite continued fraction, and we use this to conclude that the exact attractive polaron will develop into a single broad peak. Importantly, we find that the peak width scales as $T^{3/4}$ at low temperature for sufficiently weak boson-boson interactions.  We also investigate how the impurity spectral function is modified in the presence of boson-boson interactions, and in particular we argue that the width of the polaron peak will scale linearly with $T$ at low $T$ and for strong boson-impurity interactions. This is consistent with recent experiments~\cite{Yan2019}.

The paper is organized as follows. In Section~\ref{sec:hamiltonian} we describe our model for an impurity in a Bose gas, and in Section~\ref{sec:formalism} we outline the finite-temperature variational approach that we employ in this work. We discuss the universal nature of the polaron ground state in Section~\ref{sec:ground}, where we also compare with recent quantum Monte Carlo results and with experimental measurements. In Section~\ref{sec:ansatzes}, we present our calculated spectral response for the finite temperature polaron in the limit of an ideal Bose gas, and in Section~\ref{sec:origin} we explain the origin of the splitting of the attractive polaron quasiparticle branch. We investigate the effect of interactions in the medium in Section~\ref{sec:interacting-results}. Finally, we conclude in Section~\ref{sec:conclusion}.

\section{Model}\label{sec:hamiltonian}

We consider an impurity immersed in a uniform Bose gas at finite temperature. We assume that the medium is weakly interacting with $n \ab^3\ll1$, such that we can treat it within Bogoliubov theory~(see, e.g.,~Ref.~\cite{Pethick_BEC_Book}). Here, $n$ is the medium density and $\ab$ is the boson-boson $s$-wave scattering length. Measuring energy from that of the medium in the absence of the impurity, we thus have the medium-only Hamiltonian
\begin{align}
\hat{H}_0 = \sum_{\k} E_\k \num \beta \k \label{eqn:H0},
\end{align}
where the operator $\cre \beta\k$ creates a Bogoliubov excitation with momentum $\k$ and dispersion
\begin{align}
E_\k &= \sqrt{\ekb(\ekb+ 2g_{B}n_0)}.
\end{align}
Here, $\ekb=|\k|^2/2m_B \equiv k^2/2m_B$ is the free boson dispersion with $m_B$ the boson mass, while $n_0$ is the condensate density, and we define $g_B\equiv 4\pi \ab/m_B$.  The Bogoliubov creation operator $\cre \beta\k$ is related to the bare boson creation operator $\cre b\k$ by the transformation~\cite{Pethick_BEC_Book}, \beq \ann b\k = u_\k \ann \beta \k - v_\k \cre{\beta}{-\k}.  \eeq The coefficients are
\begin{subequations}
\begin{align}
u_\k &= \sqrt{\frac{1}{2}\left(\frac{\ekb+g_{B}n_0 }{E_\k} + 1 \right)}, \\
v_\k &= \sqrt{\frac{1}{2}\left(\frac{\ekb+g_{B}n_0 }{E_\k} - 1 \right)}.
\end{align}
\end{subequations}
Throughout this work, we take the boson-boson scattering length $\ab$ to be positive (even when it is taken to be infinitesimal), since attractive boson-boson interactions result in an unstable Bose gas that is prone to collapse~\cite{Dalfovo1999}. We also work in units where the reduced Planck constant $\hbar$, the Boltzmann constant $k_B$, and the volume are all set to 1.

To model the Bose polaron, we consider the total Hamiltonian
\begin{align}
\hat{H} ={}& \hat{H}_0 + \sum_\k\left(\ek\num c\k + (\epsilon_{\k,d}+\nu)\num d\k\right)\nn\\
&+ g\sqrt{n_0} \sum_\k \lba \cre d\k \ann c\k + \cre c\k \ann d\k \rba \nn \\
&+ g \sum_{\p,\k} \lba \cre d\k \ann{c}{\k-\p}\ann b\p + \cre b\p \cre{c}{\k-\p} \ann d\k \rba.
\label{eqn:hamiltonian}
\end{align}
Here, $\cre c\k$ is the impurity creation operator, and we take the impurity dispersion to be $\ek=k^2/2m$ with $m$ the impurity mass. We model the interactions between the impurity and a boson using a two-channel model~\cite{Timmermans1999fri}: This introduces a closed-channel molecule creation operator $\cre d\k$, with associated dispersion $\epsilon_{\k,d}=k^2/[2(m+m_B)]$. The interaction terms in the Hamiltonian then convert a boson and an impurity into a closed-channel molecule and vice versa, where we note that we write the interaction part of the Hamiltonian in its natural basis of bare bosonic operators, as opposed to Bogoliubov operators.

The bare parameters of the model \eqref{eqn:hamiltonian} --- the coupling strength $g$, the bare detuning between open and closed channels $\nu$, and the ultraviolet momentum cut-off $\Lambda$ above which we set the coupling to zero --- can be related to the low-energy physical parameters relevant to experiment~\cite{Bruun2004}. To this end, we calculate the two-body scattering amplitude $f(\k',\k)$ from relative momentum $\k$ to $\k'$ with $|\k|=|\k'| = k$, and compare with the standard $s$-wave low-energy expression
\beq
f(\k',\k)=-\frac{1}{\frac{1}{a}+R^*k^2+ik}. 
\eeq
In this manner, we identify the boson-impurity $s$-wave scattering length $a$ and range parameter $R^*$~\cite{Bruun2004,Levinsen2011}
\begin{align}
    a=\frac{m_r}{2\pi}\left[ \sum_\k^\Lambda \frac{1}{\ek + \ekb}-\frac{\nu}{g^2} \right]^{-1},
\end{align}
and
\begin{align}
    R^*=\frac\pi{m_r^2g^2},
\end{align}
in terms of the bare parameters of the model.  Here, $m_r=m m_B/(m+m_B)$ is the reduced mass. When $a>0$, the model describes the existence of a (vacuum) Feshbach molecule (dimer), with binding energy
\begin{equation}
\eb = \frac{(\sqrt{1+4R^*/a}-1)^2}{8 m_r R^{*2}}.
\label{eqn:dimer}
\end{equation}

We can recover the single-channel (broad resonance) model by taking the limit $R^* \to 0$. Indeed, the recent experiments~\cite{Jorgensen2016,Hu2016,Yan2019} on the Bose polaron have all been in the broad-resonance regime, where $R^* \ll |a|$. However, we choose to work with a two-channel model since we require an additional short-distance parameter to set the scale of Efimov physics~\cite{Petrov2004tbp}, as we discuss in Sec.~\ref{sec:ground}.

\subsection{Bose gas at finite temperature}

At zero temperature, the Bose gas is in the ground state and there are no Bogoliubov excitations. At finite temperature $T$, we will assume that the medium is in thermal equilibrium and hence that the Bogoliubov excitations are distributed according to the Bose-Einstein distribution function, $f_\k$. Below the critical temperature~\cite{Pethick_BEC_Book}
\beq
T_c = \frac{2\pi}{\zeta(3/2)^{2/3}} \frac{n^{2/3}}{m_B},
\label{eqn:Tc}
\eeq
where $\zeta$ is the Riemann zeta function, we have
\beq
f_\k = \frac{1}{e^{E_{\k}/T} -1}, \qquad T<T_c.
\eeq
To calculate the condensate density we use Popov theory, which extends Bogoliubov theory to finite temperatures (see, e.g., Ref.~\cite{Shi1998}). We thus find $n_0$ for a given temperature by solving the implicit equation
\beq \label{eq:popov}
n=n_0 + \frac{8 n_0}{3\sqrt{\pi}} \sqrt{n_0 \ab^3} + \sum_\k \frac{\epsilon_{\k}^B+g_B n_0}{E_\k} f_\k,
\eeq
which reduces to $n_0=n\left(1-\left(T/T_c\right)^{3/2}\right)$ in the limiting case of an ideal Bose gas.  Popov theory is known to fail close to the critical temperature~\cite{Shi1998}, since it assumes that \beq \frac{\abs{T-T_c}}{T_c} \gg n^{1/3}\ab.  \eeq For small $n^{1/3}\ab$ this excludes a narrow region around the critical temperature.

Above the critical temperature, we have $n_0=0$ and Popov theory reduces to that of an ideal Bose gas. In this regime, we have $E_\k=\ekb$, $u_\k=1$, $v_\k=0$, and $\hat b_\k=\hat \beta_\k$. The Bose-Einstein distribution reduces to
\beq
f_\k = \frac{1}{e^{(\ekb - \mu)/T} -1}, \qquad T>T_c,
\eeq
where the chemical potential $\mu$ is found by solving
\beq
n=\int_0^\infty d\epsilon \frac{m_B^{3/2}}{2^{1/2} \pi^2}\frac{\epsilon^{1/2}}{e^{( \epsilon-\mu)/T}-1}
\eeq
for a given $n$ and $T$.

\section{Variational approach}\label{sec:formalism}

The framework for our investigation of the Bose polaron at finite temperature is the variational approach for impurity dynamics first developed in Ref.~\cite{Liu2019} in the context of Fermi polarons. A variational method based on including only one excitation of the medium was first used to obtain the zero-temperature ground-state properties of the Fermi polaron~\cite{Chevy2006} as well as the repulsive branch~\cite{Cui2010,Massignan2011}, and it has been demonstrated that such an approach is equivalent to a diagrammatic formulation~\cite{Combescot2007}.  We also note that related zero-temperature variational approaches have been used to obtain the energy spectrum of the Bose polaron and compare against experiment~\cite{Jorgensen2016} and to investigate the impact of Efimov physics on the Bose polaron~\cite{Levinsen2015,Yoshida2018}.  While such variational methods have primarily been used in the context of polarons in ultracold gases, they can be generalized to other systems.

For completeness, in the following we outline the variational approach of Ref.~\cite{Liu2019}. The key idea that distinguishes this method from the zero-temperature variational approach developed in Ref.~\cite{Parish2016} is to work with an impurity operator in the Heisenberg picture rather than with wave functions in the \sch picture. This allows one to separate the thermal average over medium states from the dynamics of the impurity. We therefore consider the impurity annihilation operator within the Heisenberg picture
\begin{align}
    \hat c_\0(t)=e^{i\hat Ht}\hat c_\0e^{-i\hat Ht}.
    \label{eq:Heisop}
\end{align}
For simplicity, we specialize to the case of an initially non-interacting impurity at rest, and we take $\hat H$ to be time-independent. Both of these conditions can be relaxed --- see Ref.~\cite{Liu2019}.

Since the dynamics is generally not exactly solvable, we introduce an approximate impurity annihilation operator $\hat{\mathbf c}_\0(t)$. Unlike the exact operator, this does not satisfy the Heisenberg equation of motion. Instead, we define an error operator
\begin{align}
    \hat \epsilon(t)=i\del_t\hat{\mathbf c}_\0(t)-[\hat{\mathbf c}_\0(t),\hat H],
\end{align}
that would vanish if $\hat{\mathbf c}_\0(t)$ were the exact operator. To arrive at a scalar quantity, we then take the expectation value of the Hermitian operator $\ann{\epsilon}{}(t)\cre{\epsilon}{}(t)$,
\beq
\Delta(t) = {\rm Tr}[\hat{\rho_0} \ann{\epsilon}{}(t)\cre{\epsilon}{}(t) ],
\label{eq:Delta}
\eeq
Here the trace is over all realizations of the medium in the absence of the impurity, and we work in Fock space such that the impurity operator can act directly on a medium state. Thus, $\Delta(t)$ can be viewed as quantifying the error accumulated in the approximate impurity operator for a particular realization of the medium. While the method can in principle be used for any mixed state, we will consider only a thermal state at temperature $T$, for which the medium density matrix is $\hat \rho_0=e^{-\hat{H}_0/T} / {\rm Tr}(e^{-\hat{H}_0/T})$. Note that since the trace is over medium-only states, we have ${\rm Tr}[e^{-\hat{H}_0/T}\cdots]={\rm Tr}[e^{-\hat{H}/T}\cdots]$. In the following we use the notation $\langle \cdots \rangle\equiv {\rm Tr}[\hat \rho_0\cdots]$ to indicate a thermal average.

Following Ref.~\cite{Liu2019}, our ansatz for the impurity annihilation operator is a linear combination of operators of the general form
\beq
\hat{\mathbf c}_\0(t) = \sum_j \eta_j(t) \ann Oj,
\eeq
where $\eta_j(t)$ are time-dependent variational parameters and $\ann Oj$ are time-independent operators that form the basis for our ansatz. These operators are unique products of impurity and medium operators, and importantly they all contain precisely one impurity: $\langle \ann Oj \hat N_{\rm imp}\cre Oj\rangle=1$ with $\hat N_{\rm imp}\equiv \sum_\k[\num c\k + \num d\k]$. Furthermore, we choose the operators to be orthogonal under the thermal average, i.e., $\langle \ann Oj \cre Ol \rangle = 0$ if $j\ne l$.

To derive the equations of motion for the coefficients $\eta_j(t)$, we impose the minimization condition $\frac{\del \Delta}{\del (d\eta_j^*/dt)}=0$ on the error quantity $\Delta(t)$ in Eq.~\eqref{eq:Delta}. This results in
\beq
i \frac{d\eta_j}{dt} \langle \ann Oj \cre Oj \rangle = \sum_i \eta_i(t) \langle [\ann Oi, \hat{H}] \cre Oj \rangle.
\label{eqn:TBM-time-dependent}
\eeq The time evolution governed by this equation conserves probability~\cite{Liu2019}; that is, $\langle \hat{\mathbf c}_\0(t) \hat{\mathbf c}^\dag_\0(t) \rangle$ is constant.

Specializing to the stationary variational impurity operators, these have coefficients of the form $\eta_j(t)=\eta_j e^{-iEt}$. This gives the time-independent version of Eq.~\eqref{eqn:TBM-time-dependent},
\beq
E \eta_j \langle \ann Oj \cre Oj \rangle = \sum_i \eta_i \langle [\ann Oi, \hat{H}] \cre Oj \rangle,
\label{eqn:TBM}
\eeq
which is a system of linear equations for the coefficients $\eta_j$ (or, equivalently, $\eta^*_j$). From the resulting eigenenergies $E_l$ and associated eigenvectors $\eta_j^{(l)}$, we then construct approximate stationary impurity operators
\begin{align}
    \hat \phi_l=\sum_j\eta_j^{(l)}\ann Oj,
    \label{eq:stationary}
\end{align}
where we normalize the eigenvector $\eta_j^{(l)}$ such that the stationary operators are orthonormal under the thermal average:
\begin{align}
    \langle \hat \phi_l\hat\phi^\dag_m\rangle=\delta_{lm}.
\end{align}
Since the operators $\hat O_j$ form a complete basis for $\hat {\mathbf c}_\0$, so do the stationary operators. We can therefore construct the impurity operator as
\begin{align}
    \hat {\mathbf c}_\0(t)=\sum_l\kappa_l\hat \phi_le^{-iE_lt},
\end{align}
with $\kappa_l$ an expansion coefficient. From the orthogonality of the stationary operators, we immediately see that $\kappa_l=\langle \hat {\mathbf c}_\0(0)\hat \phi_l^\dag\rangle=\langle \hat {c}_\0\hat \phi_l^\dag\rangle$, where in the last step we took the impurity to be initially non-interacting. Therefore, we finally obtain
\begin{align}
    \hat {\mathbf c}_\0(t)=\sum_l\langle \hat { c}_\0\hat \phi_l^\dag\rangle \hat \phi_l e^{-iE_lt}.
    \label{eq:ct}
\end{align}
This expression can be used to construct the experimental observables of interest.

In general, we determine the types of basis operators that we consider in our variational ansatz by taking the initial operator $\hat c_\0$ and commuting it a number of times with the Hamiltonian.
These operators correspond to the lowest-order terms that would be generated if one were to consider the short-time behavior of the impurity operator in the Heisenberg picture, Eq.~\eqref{eq:Heisop}. 
However, we emphasize that since our approach is variational, it can be extended beyond the perturbative regime.

\subsection{Impurity spectral response}

An experimental protocol of particular interest is that of inverse (or injection) radiofrequency spectroscopy, where an initially non-interacting impurity is suddenly introduced into the medium. Within linear response, the probability of this process is proportional to the impurity spectral function~\cite{mahan},
\begin{align}
    A(\omega)={\rm Re}\int_0^\infty \frac{dt}\pi\, e^{i\omega t}\langle \hat c_\0(t)\hat c_\0^\dag\rangle,
    \label{eq:Aomega}
\end{align}
where, for simplicity, we again consider an impurity at rest and we measure the impurity energy from that of the non-interacting initial state. Within our approximation, we insert Eq.~\eqref{eq:ct} in Eq.~\eqref{eq:Aomega} to find
\begin{align}
    A(\omega)=\sum_l \left|\langle \hat {c}_\0 \hat \phi_l^\dag\rangle\right|^2\delta(\omega-E_l).
\end{align}

Because integrals are discrete within our numerical calculations, in practice we calculate a finite number of discrete energy eigenvalues. To recover continua, to enhance visibility of narrow features, and to capture the effects of a finite linewidth in experiments, we convolve the spectral function with a Gaussian of width $\sigma$ to obtain the broadened spectral function
\beq
I(E) = \sum_l \left|\langle \hat {c}_\0 \hat \phi_l^\dag\rangle\right|^2 \frac{1}{\sqrt{2\pi} \sigma}e^{-(E-E_{l})^2/2\sigma^2}.
\label{eqn:spec-broad}
\eeq
This broadened spectral function is what we show in most of this paper. We choose the width $\sigma$ to be comparable to the Fourier broadening in experiment so that our results provide a direct comparison to the experimentally measured energy spectra~\cite{Hu2016,Jorgensen2016}. 

\section{Universality of the polaron ground state} \label{sec:ground}

We start by considering the Bose polaron at $T=0$. In this case, we employ an approximate impurity creation operator of the form
\begin{align}
    \hat{\mathbf c}^\dag_\0=&\alpha_0 \hat c^\dag_\0+\sum_\k \alpha_\k \hat c^\dag_{-\k}\hat\beta^\dag_\k+\frac12\sum_{\k_1\k_2}\alpha_{\k_1\k_2}\hat c^\dag_{-\k_1-\k_2}\hat\beta^\dag_{\k_1}\hat\beta^\dag_{\k_2} \nn \\
    &+\gamma_0\hat d^\dag_\0+\sum_\k \gamma_\k\hat d^\dag_{-\k}\hat \beta^\dag_\k,
    \label{eq:variation0}
\end{align}
where we suppress the explicit time dependence, noting that throughout this paper we will be concerned with the stationary operators introduced in Eq.~\eqref{eq:stationary}. This impurity operator explicitly contains up to three-body correlations (the impurity plus two Bogoliubov excitations), and the general form of the operator can be obtained by commuting the bare impurity operator four times with the Hamiltonian, while keeping only excitations of the condensate. The first line of Eq.~\eqref{eq:variation0} consists of terms with the impurity and 0, 1, or 2 Bogoliubov operators, while in the second line we have the closed-channel molecule by itself or with a single Bogoliubov operator.  Note that we have the symmetry $\alpha_{\k_1\k_2} = \alpha_{\k_2\k_1}$ since the Bogoliubov excitations correspond to identical bosons.  It is straightforward to see how to extend the number of terms in the variational ansatz; however the computational complexity of the variational approach increases rapidly with the number of operators.

At zero temperature, the variational approach outlined in Sec.~\ref{sec:formalism} above reduces to the variational wave function approach developed in Ref.~\cite{Parish2016} in the context of the Fermi polaron, and applied to the Bose polaron in Ref.~\cite{Jorgensen2016}. In Ref.~\cite{Jorgensen2016}, the impurity spectral function was evaluated variationally and was seen to match very well with the experimentally measured spectral response. To clearly see the connection between the two approaches note that, at $T=0$, we can convert the impurity operator to a variational wave function
\begin{align}
    \ket{\Psi(t)}=\hat{\mathbf c}^\dag_\0(t)\ket{\Phi},
\end{align}
where $\ket{\Phi}$ is the Bose gas ground state, and $\hat{\mathbf c}^\dag_\0(t)$ is given by Eq.~\eqref{eq:variation0}. Apart from the time-dependence, this is precisely the variational ansatz considered in Ref.~\cite{Jorgensen2016}, and the stationary solutions precisely correspond to the approximate energy-eigenstates considered in Ref.~\cite{Jorgensen2016}. Indeed, we find that the set of linear equations~\eqref{eqn:TBM} for the stationary coefficients,
\begin{subequations}
\label{eq:2bogT0}
	\begin{align}
	E\alpha_0 = & \ g \sqrt{n_0} \gamma_0 - g \sum_\k v_\k \gamma_\k \\
	E\gamma_0 = & \  \nu \gamma_0 + g \sqrt{n_0} \alpha_0 + g \sum_{\k} u_\k \alpha_{\k} \\
    E\alpha_\k = & \ (\ek + E_\k) \alpha_\k+g u_\k \gamma_0 +g\sqrt{n_0}\gamma_\k\\ \nn
    E\gamma_\k = & \ (\epsilon_{\k,d}+\nu+E_\k)\gamma_\k-gv_\k \alpha_0 + g\sqrt{n_0}\alpha_\k \\
    & \ +g\sum_{\k'} u_{\k'}\alpha_{\k\k'}\\ \nn
    E\alpha_{\k_1\k_2} = & \ (E_{\k_1}+E_{\k_2}+\epsilon_{\k_1+\k_2})\alpha_{\k_1\k_2} \\
    & \ +g(u_{\k_2}\gamma_{\k_1}+u_{\k_1}\gamma_{\k_2}),
	\end{align}
\end{subequations}
exactly matches that considered in Ref.~\cite{Jorgensen2016} (see also Ref.~\cite{Levinsen2015} where such a wave function was first used to obtain the polaron ground state). Hence, the variational approaches based on wave functions~\cite{Parish2016} and operators~\cite{Liu2019} are equivalent at zero temperature.

\begin{figure}
    \centering
    \includegraphics{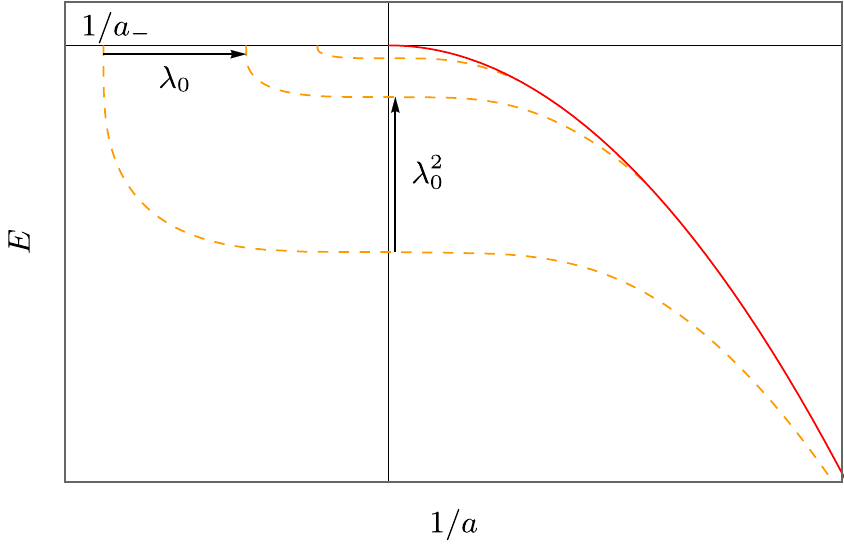}
    \caption{Schematic depiction of the Efimov trimer spectrum. The Efimov trimers are shown as dashed lines and the dimer state as a solid line. The spectrum is invariant under a scaling transformation that takes $1/a\to1/(\lambda_0a)$ and $E\to E/\lambda_0^2$. 
    }
    \label{fig:EfimovSketch}
\end{figure}

As noted above, the variational ansatz~\eqref{eq:variation0} explicitly includes three-body correlations involving the impurity plus two bosons. This has two distinct advantages~\cite{Levinsen2015}: First, this is the minimal complexity that allows one to correctly recover the weak-coupling perturbation theory results to second~\cite{Novikov2009} and third~\cite{Christensen2015} order in the impurity-boson scattering length.  In other words, it is the minimal description that allows one to accurately go beyond mean field theory.  Second, including three-body correlations allows one to capture the connection between polaron physics and non-trivial few-body physics such as Efimov trimers~\cite{Efimov1970}, illustrated in Fig.~\ref{fig:EfimovSketch}. Indeed, we immediately see that in the limit of zero density, 
the linear set of equations~\eqref{eq:2bogT0} separates into a 1-body sector [Eq.~(\ref{eq:2bogT0}a)], a two-body sector [Eqs.~(\ref{eq:2bogT0}b-c)] and a three-body sector [Eqs.~(\ref{eq:2bogT0}d-e)], where the latter precisely includes all three-body bound states~\cite{Levinsen2015}. We emphasize that this property holds even for the \textit{exact} impurity operator with an infinite number of terms, and thus the low-density limit should always recover the few-body spectrum.

The peculiar few-body phenomenon of Efimov physics is a particularly intriguing aspect of the Bose polaron~\cite{Levinsen2015}. We therefore now briefly discuss the salient points of Efimov physics --- we refer the reader to Ref.~\cite{naidon2017} for a recent review. Efimov's original prediction was that three identical bosons with near resonant two-body interactions can form trimer bound states with a spectrum that features a discrete scaling symmetry~\cite{Efimov1970}, as shown in Fig.~\ref{fig:EfimovSketch}. That is, in the limit of a very large boson-boson scattering length $\ab$ compared with the typical range of the potential, each trimer in the spectrum is exactly reproduced under the rescaling of the scattering length $\ab\to \lambda_0^l \ab$ and energy $E\to E\lambda_0^{-2l}$, with $l$ an integer and scaling parameter $\lambda_0\simeq22.7$. Remarkably, these trimers can exist even in the absence of a two-body bound state.  Such Efimov trimers were first experimentally observed in an ultracold Bose gas of Cs atoms~\cite{Kraemer2006}.

Since the first prediction of Efimov trimers, the concept has been extended to the scenario of a single particle (in the present context, we call this an impurity) that strongly interacts with two identical bosons~\cite{Efimov1973}. Here, one essentially finds the same physics, where the impurity-boson scattering length $a$ is now the parameter governing the Efimov spectrum in Fig.~\ref{fig:EfimovSketch}.  Most recently, it was shown that this impurity problem supports the formation of even larger clusters such as tetramers~\cite{Blume2014,Yoshida2018}, and even pentamers and hexamers~\cite{Blume2019}.

In the impurity scenario, the Efimov scaling parameter $\lambda_0$ depends strongly on the mass ratio. For instance, in the case of equal masses $m=m_B$ and assuming $|a|\gg \ab$, we have $\lambda_0\simeq1986.1$~\cite{Efimov1973}. Furthermore, $\lambda_0$ increases rapidly with $m/m_B$ until it diverges as $m/m_B\to\infty$~\cite{Efimov1973}. This large separation of scales between successive trimers is reflected in the ratio of the typical length scale of the short-range interaction (which in our case is $\lesssim R^*$), and the length scale associated with the ground-state trimer.  For instance, for equal masses, the critical scattering length at which the ground-state trimer unbinds into the continuum is $a_-=-4934R^*$~\cite{Yoshida2018}. This was shown in Ref.~\cite{Yoshida2018} to imply that the few-body physics is nearly model independent: No matter how one introduces an ultraviolet cutoff, the few-body spectrum depends only on $a_-$, and the details of the short-range physics only cause minute corrections. In particular, the ratio between the ground-state trimer and tetramer energies at unitarity was found to be essentially universal for three different models~\footnote{Results from a recent work imply that this few-body universality requires there to be an effective short-range repulsion between bosons~\cite{Blume2019}.}.

This universality also extends to the many-body case of the Bose polaron, since the polaron ground-state energy for the equal-mass case was recently shown to be a universal function of the dimensionless three-body parameter $n^{1/3} |a_-|$ in the unitary limit $1/a=0$~\cite{Yoshida2018}.  However, a remaining question is whether the Bose polaron remains universally dependent on the Efimov scale when the scattering length is varied away from unitarity.

\begin{figure}
    \centering
    \includegraphics{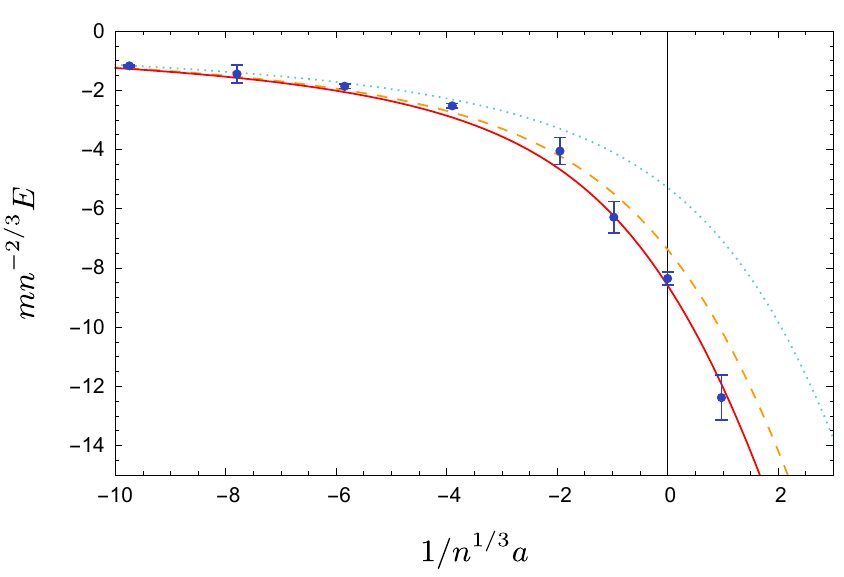}
    \caption{Ground-state polaron energy for the equal-mass case calculated within our variational approach (lines) and in a recent QMC study~\cite{Pena-Ardila2019} (symbols). We show the results of dressing the impurity by up to 1, 2, or 3 excitations of the Bose gas (dotted, dashed, and solid lines, respectively), and we take $n^{1/3}R^*=0.015$ such that we have three-body parameter 
    $n^{1/3}|a_-| \simeq 70$, which is the same as in the QMC (see text).
  }
    \label{fig:QMCcompare}
\end{figure}

\begin{figure}
    \centering
    \includegraphics{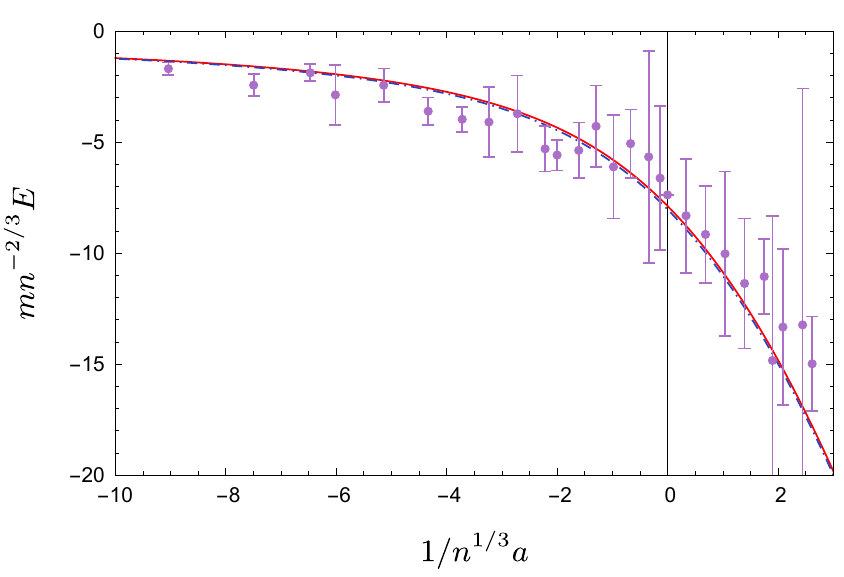}
    \caption{Ground-state polaron energy for the equal-mass case calculated within our variational approach (lines) using 3 Bogoliubov excitations, compared with the experimentally measured ground-state energy (symbols)~\cite{Jorgensen2016} (see also Ref.~\cite{Pena-Ardila2019}). We take $n=4\times10^{14}{\rm cm}^{-3}$, $R^*=60a_0$, and $\ab=9a_0$ as in the experiment, which corresponds to $n^{1/3} R^* \simeq 0.02$ and $n^{1/3}\ab \simeq 3.5 \times 10^{-3}$  (solid line). We also include the results where we take the ideal gas limit $\ab \to 0$ (dotted line).}
    \label{fig:EXPcompare}
\end{figure}

To address this question, we have calculated the polaron ground-state energy as a function of $1/n^{1/3}a$ and compared it with the results from a recent QMC study, as displayed in Fig.~\ref{fig:QMCcompare}.  Here, unlike in Eq.~\eqref{eq:variation0}, we have included up to three excitations of the medium, i.e., \textit{four}-body correlations. We do not write this ansatz or its associated linear equations here, but refer the reader to Ref.~\cite{Yoshida2018} where these were first presented.  In the QMC calculations~\cite{Pena-Ardila2019}, the boson-boson repulsion was modelled as a hard-sphere potential, corresponding to $n^{1/3}\ab \simeq 3.5 \times 10^{-3}$, which we convert into an Efimov three-body parameter using the relationship $a_- \simeq -2 \times 10^4 \ab$ for the hard-core boson model with $m=m_B$~\cite{Yoshida2018}.

Referring to Fig.~\ref{fig:QMCcompare}, we see that the polaron energy from the variational approach converges to the QMC result as we increase the number of boson excitations in the ansatz. Note that we take $\ab \to 0$ in our variational calculations, since this is essentially indistinguishable from using the tiny $n^{1/3}\ab$ in the QMC calculations (see, also, Fig.~\ref{fig:EXPcompare}).  Crucially, we find that including three excitations yields an excellent agreement across the whole range of scattering lengths, while including two excitations (three-body correlations) is already accurate for $1/n^{1/3}a \lesssim -1$. This is even more remarkable given that the ground-state energy diverges when $R^* \to 0$~\cite{Yoshida2018}.  Thus, this explicitly demonstrates the accuracy and universality of our approach, as well as showing that Efimov physics is crucial for correctly predicting the ground-state energy close to resonance.

To further expose the universal role of Efimov physics in the Bose polaron, we also compare our calculated ground-state energy with that recently extracted~\cite{Pena-Ardila2019} from experimental measurements~\cite{Jorgensen2016}, as shown in Fig.~\ref{fig:EXPcompare}.  Once again, we find very good agreement across the whole region around unitarity when we employ a variational ansatz with three Bogoliubov excitations.  Here we use the range parameter $n^{1/3}R^* \simeq 0.02$ taken from experiment~\cite{Jorgensen2016}, which yields a slightly larger three-body parameter $n^{1/3}|a_-|$ than in Fig.~\ref{fig:QMCcompare}.  Had we used the same parameters as in the QMC calculations, then the fit with the experiment would have been noticeably worse, e.g., the calculated energy at unitarity would be shifted downwards by $7$\%.  On the other hand, we see that the effect of the boson-boson interactions is essentially negligible when $n^{1/3}\ab \ll 1$.  Therefore, the key length scale in the Aarhus experiment~\cite{Jorgensen2016} is the Efimov scale rather than the low-energy boson-boson scattering length.

\section{Finite-temperature Bose polaron}\label{sec:ansatzes}

To elucidate the finite-temperature behavior of the Bose polaron, we first take the limit of an ideal Bose gas, where $\ab \to 0^+$.  This simplifies the analysis since it allows us to exactly model the bosonic medium at finite temperature and it reduces the number of length scales in the problem.  We investigate the effect of boson-boson interactions on the polaron energy spectrum in Sec.~\ref{sec:interacting-results}.

In the following, we consider two ansatzes for the approximate impurity operator that include up to two- and three-body correlations, respectively. The ``two-body'' ansatz is equivalent to the extended ladder approximation from Ref.~\cite{Guenther2018}, while the more sophisticated ``three-body'' ansatz includes scattering processes that go beyond previous work and allows us to capture Efimov physics at arbitrary temperature.

\subsection{Polaron with two-body correlations}\label{sec:two-body-results}

To generate the terms in the two-body ansatz, we first take the limit $\ab \to 0$ and then commute the bare impurity operator twice with the Hamiltonian~\eqref{eqn:hamiltonian}. This yields the variational operator 
\begin{align} \nn
\hat{\mathbf{c}}^\dag_\0 =&{} \alpha_0 \cre c\0 + \gamma_0 \cre d\0 + \sum_\q \gamma^\q \cre{d}{\q} \ann b\q + \sum_\q \alpha^\q \cre{c}{\q} \ann b\q\\ \label{eqn:ansatz-two-body}
&+ \sum_\k \alpha_\k \cre{c}{-\k} \cre b\k + \sum_{\k, \q} \alpha^\q_\k \cre{c}{\q-\k} \cre b\k \ann b\q, 
\end{align}
which contains all two-body correlations for the polaron, i.e., all scattering processes involving the impurity and up to one boson.  Like in Eq.~\eqref{eq:variation0}, the use of $\alpha$ or $\gamma$ denote terms with a bare impurity or a closed-channel dimer, respectively.  However, in contrast to the zero-temperature case, Eq.~\eqref{eqn:ansatz-two-body} also includes terms with 1 hole excitation at finite momentum (denoted by coefficients with a superscript), corresponding to a boson being removed from the thermal cloud. Such a boson can either be scattered into the condensate or into another finite-momentum state.

\begin{figure*}
    \centering
    \includegraphics{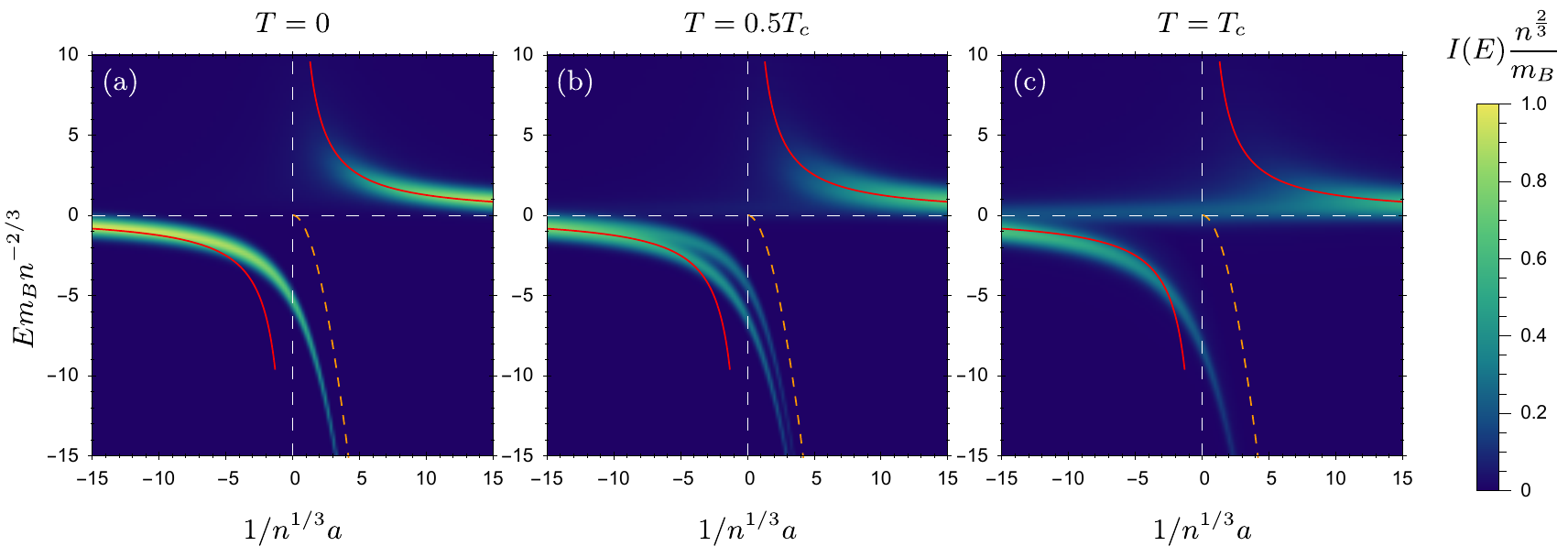}
    \caption{Impurity spectral functions obtained from the two-body ansatz \eqref{eqn:ansatz-two-body} 
      for different temperatures (a) $T=0$, (b) $T=0.5T_c$ and (c) $T=T_c$. We use a broadening of $\sigma=0.4 n^{2/3}/m_B$, which is comparable to the Fourier broadening in the Aarhus experiment~\cite{Jorgensen2016}, and we take $n^{1/3} R^*=0.02$ and $m=m_B$. The solid red lines are the mean-field energy \eqref{eqn:meanfield}, while the dashed orange lines are the dimer energy $-\eb$.  }
    \label{fig:two-body}
\end{figure*}

The system of linear equations~\eqref{eqn:TBM} for the two-body ansatz is
\begin{subequations} \label{eqn:two-body-equations}
	\begin{align} 
	E\alpha_0 &= g \sqrt{n_0} \gamma_0 + g \sum_{\q} f_{\q} \gamma^{\q} \\
	E\gamma_0 &= \nu \gamma_0 + g \sqrt{n_0} \alpha_0  + g \sum_{\k} (1+f_{\k}) \alpha_{\k} \\
	E\gamma^\q &=  (\epsilon_{\q,d}+\nu-\eqb) \gamma^{\q} + g \alpha_0 + g \sqrt{n_0} \alpha^\q \nn \\
	&\quad + g \sum_{\k} (1+f_{\k}) \alpha^\q_{\k}  \\
	E\alpha^\q &=  (\eq - \eqb)\alpha^\q + g\sqrt{n_0} \gamma^\q \\
	E\alpha_\k &= (\ek + \ekb) \alpha_\k + g \gamma_0 \\
	E\alpha^\q_\k &=  (\epsilon_{\q-\k} + \ekb - \eqb) \alpha^\q_\k + g \gamma^\q .
	\end{align}
\end{subequations}
We obtain the polaron energy spectrum by expressing this set of equations as a symmetric matrix and then solving for the eigenvectors and eigenvalues --- see Appendix~\ref{sec:numerics} for further details.  We then calculate the broadened spectral function in Eq.~\eqref{eqn:spec-broad}, which corresponds to
\beq
I(E) = \sum_l \big|\alpha^{(l)}_0\big|^2 \frac{1}{\sqrt{2\pi} \sigma}e^{-(E-E_{l})^2/2\sigma^2},
\eeq
where $l$ indexes the different stationary operators, with corresponding energy eigenvalues $E_l$.

Before turning to our results for the spectrum, it is first instructive to see how our approach is related to standard finite-temperature Green's function approaches~\cite{Fetter}. Rearranging Eq.~\eqref{eqn:two-body-equations}, we can eliminate the coefficients to obtain an implicit equation for the impurity energy
\begin{align} \nn
    E = \ & n_0 \left[\frac{m_r}{2\pi a} 
    + \sum_\k \left(\frac{1+f_\k}{-E+\ek + \ekb} -\frac1{\ek+\ekb} \right) \right]^{-1} \\ \nn
    & + \sum_\q f_\q \left[\frac{m_r}{2\pi a}
    - \frac{n_0}{E - \eq + \eqb} \right. \\ \label{eq:extTmat}
   & \left. + \sum_\k \left(\frac{1+f_\k}{-E+\epsilon_{\q-\k} + \ekb - \eqb} -\frac1{\ek+\ekb} \right) \right]^{-1} .
\end{align}
Here, we have taken the limit $R^* \to 0$ for simplicity since the two-body ansatz does not contain Efimov physics and thus we do not require an additional short-distance length scale.  Equation~\eqref{eq:extTmat} is equivalent to the pole condition of the impurity Green's function, $E = \Sigma(E)$, where $\Sigma(E)$ is the impurity self-energy within the extended ladder approximation introduced in Ref.~\cite{Guenther2018}. The first term in the self energy is the energy shift due to interactions with the condensate only, while the second term involves scattering of the impurity with the thermal cloud and it thus disappears at zero temperature.  In Ref.~\cite{Guenther2018}, the additional condensate term $n_0/(E - \eq + \eqb)$ had to be introduced by hand as an extension to the ladder diagrams, whereas we see here that it naturally appears in the variational approach and it is linked to the $\alpha^\q$ term of the two-body ansatz in Eq.~\eqref{eqn:ansatz-two-body}.

Figure~\ref{fig:two-body} displays the impurity spectral function calculated within the variational approach for different interaction strengths $1/n^{1/3}a$ and temperatures $T\leq T_c$. 
Here we consider the equal-mass case and we take $n^{1/3}R^*=0.02$, since this is approximately the range parameter in the Aarhus experiment~\cite{Jorgensen2016}. However, since the two-body ansatz does not contain any Efimov physics, the length scale $R^*$ only has a small effect on the spectrum when $n^{1/3}R^* \ll 1$, so it can essentially be regarded as zero for the plotted energy range in Fig.~\ref{fig:two-body}.

At zero temperature, we obtain attractive and repulsive polaron branches at negative and positive energies, respectively, which is consistent with previous theoretical works involving the ladder approximation~\cite{Li2014,Rath2013}.  For weak interactions $n^{1/3} |a| \ll 1$, the energy shifts of these branches are given by the mean-field result
\begin{equation}
E_{MF} = \frac{2\pi n a}{m_r}.
\label{eqn:meanfield}
\end{equation}
In general, the zero-temperature energy spectrum only depends on the reduced mass $m_r$ within the two-body ansatz, since it only contains impurity-boson scattering at zero center-of-mass momentum. Thus the spectrum in Fig.~\ref{fig:two-body}(a) has the same form for any impurity-boson mass ratio.

The attractive polaron in Fig.~\ref{fig:two-body}(a) corresponds to the ground state, and it smoothly evolves into the two-body dimer state as $1/n^{1/3}a \to \infty$, rather than undergoing a sharp transition like in the case of the Fermi polaron~\cite{Prokofev2008}. Note that the attractive Bose polaron is actually expected to evolve into higher body bound states, but these are not present in the two-body ansatz. The repulsive polaron branch, by contrast, corresponds to a collection of excited eigenstates and thus manifests as a broadened peak. The exact spectrum should also feature a continuum of states between the polaron branches, but this is not captured by the two-body ansatz for energies $E<0$.

Progressing to finite temperatures, several interesting features emerge that go beyond simple thermal decoherence. In Fig.~\ref{fig:two-body}(b), where we plot the spectrum for $T=0.5T_c$, we see that the attractive polaron splits into two branches, like what was observed in Ref.~\cite{Guenther2018}. This splitting persists up until the critical temperature $T=T_c$, at which point the upper attractive branch disappears and only the lower branch remains, as shown in Fig.~\ref{fig:two-body}(c).  We also observe the emergence of a broad zero-energy peak for all $1/n^{1/3}a$ at finite temperature, which becomes most prominent for $T\lesssim T_c$.

\begin{figure}
    \centering
    \includegraphics{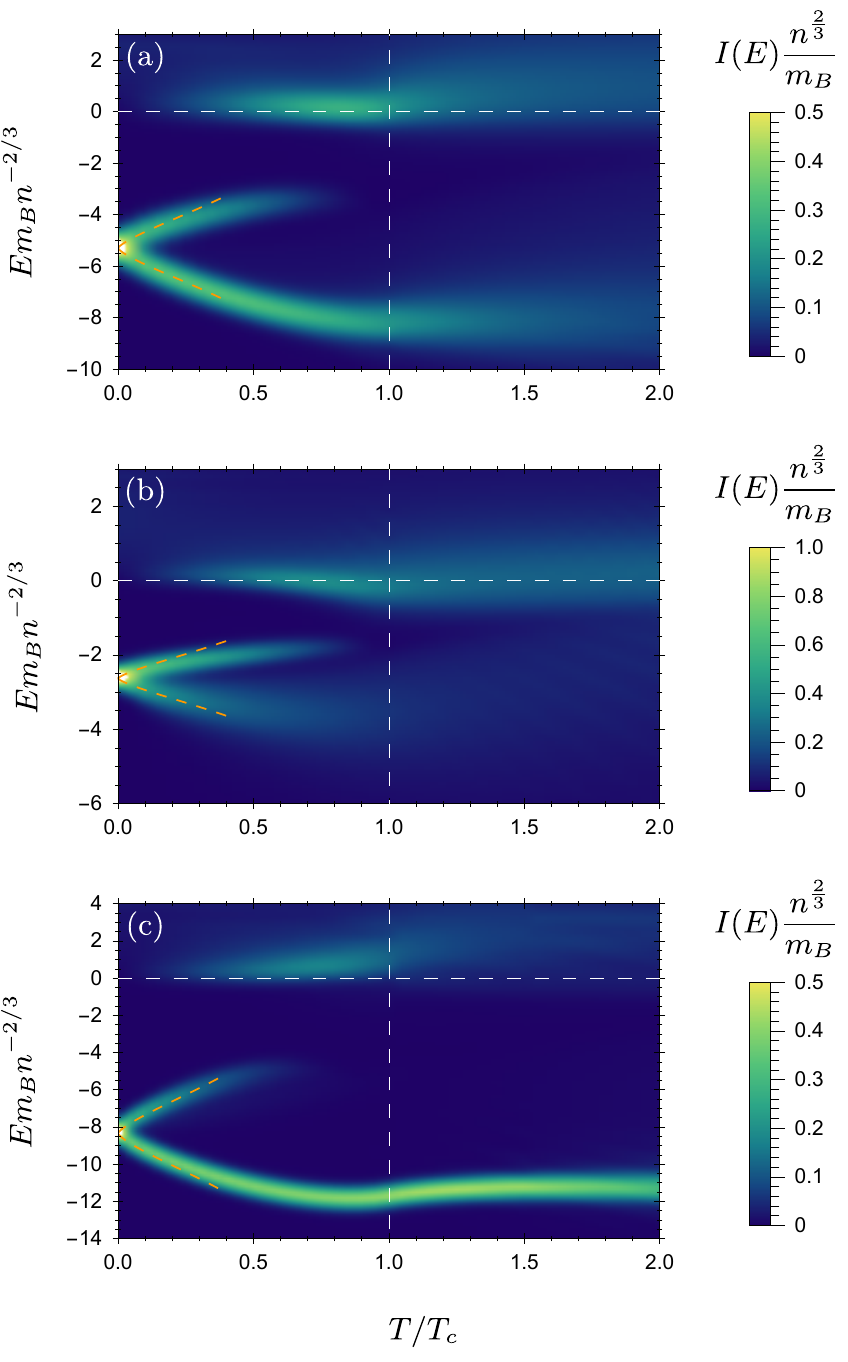}
    \caption{Impurity spectra obtained from the two-body ansatz \eqref{eqn:ansatz-two-body} at unitarity 
    $1/a=0$ and plotted with respect to temperature. (a) The equal-mass case,  $m=m_B$,  with broadening $\sigma=0.4 n^{2/3}/m_B$. 
    (b) The fixed impurity, $m\to\infty$, with smaller broadening $\sigma=0.2 n^{2/3}/m_B$ due to the smaller polaron energy scale. 
    (c) The light impurity with broadening $\sigma=0.4 n^{2/3}/m_B$ and $m/m_B=40/87$, corresponding to the mass ratio in the JILA experiment~\cite{Hu2016}. In all plots, we take $n^{1/3}R^*=0.02$. The orange dashed lines are the predicted energies  \eqref{eq:DE2body} from a low-temperature analysis.
    }
    \label{fig:two-body-temperature}
\end{figure}

To better illustrate the temperature dependence of these features, we plot the spectrum versus temperature at unitarity ($1/a=0$) in Fig. \ref{fig:two-body-temperature}.  The equal-mass spectral function plotted in Fig.~\ref{fig:two-body-temperature}(a) is consistent with that predicted in Ref.~\cite{Guenther2018}.  In addition to the equal-mass case, we consider the spectra for an infinitely heavy impurity ($m \to \infty$) and a light impurity with $m/m_B = 40/87$, which corresponds to the impurity-boson mass ratio in the JILA experiment~\cite{Hu2016}.  For all mass ratios, we see that the ground-state polaron evenly splits into two branches as soon as the temperature is raised from zero.  The size of this splitting is proportional to the ground-state polaron energy, as we show formally in Sec.~\ref{sec:origin}.  Moreover, the spectral weight of the upper attractive branch always decreases to zero as the temperature approaches the critical temperature.

However, the impurity-boson mass ratio does affect the relative broadening of the attractive branches at finite temperature.  In particular, we see that the lower attractive branch broadens substantially for the infinitely heavy impurity in Fig.~\ref{fig:two-body-temperature}(b), while it remains relatively narrow for the light impurity case, even above the critical temperature, as shown in Fig.~\ref{fig:two-body-temperature}(c).  This is derived from the fact that the zero-temperature attractive polaron energy is $E_{\rm att} \sim n^{2/3}/m_r$ at unitarity, whereas the critical temperature $T_c$ in Eq.~\eqref{eqn:Tc} only depends on $m_B$. Therefore, we have $T_c/E_{\rm att} \sim m_r/m_B = m/(m+m_B)$. This implies that the relative temperature at $T_c$ is smaller for lighter impurities and thus the thermal broadening at a given $T/T_c$ in Fig.~\ref{fig:two-body-temperature}(c) should be decreased by a factor of 1.6 compared to Fig.~\ref{fig:two-body-temperature}(a).  Similarly, the polaron broadening for the fixed impurity at a given $T/T_c$ in Fig.~\ref{fig:two-body-temperature}(b) is increased by a factor of two compared to the equal-mass case.

The other spectral feature that emerges at finite temperature is a peak around zero energy, which increases in amplitude as the temperature increases up to the critical temperature. Unlike the attractive polaron branches, which consist of a single dominant eigenstate with a large overlap with the non-interacting impurity, this zero-energy feature consists of many eigenstates with a small overlap with the non-interacting impurity.  Specifically, for $T<T_c$, the majority of the weight of each state within the zero-energy peak is contained in the $\alpha_\k$-term of Eq.~\eqref{eqn:ansatz-two-body} with momentum $|\k|$ close to zero. Physically, this means that the zero-energy peak primarily involves scattering processes where bosons are scattered out of the condensate and into low-momentum states of the thermal cloud. The enhancement of such scattering processes is connected to the singular nature of the Bose distribution $f_\k$ at low momentum when $T<T_c$, as we illustrate in Sec.~\ref{sec:interacting-results}.

In the high-temperature limit $T \to \infty$, our two-body ansatz becomes equivalent to the second-order term in the virial expansion for the Bose polaron~\cite{Sun2017}. One can see this by expanding Eq.~\eqref{eq:extTmat} to lowest order in the fugacity $e^{\mu/T}$.  With increasing temperature, we approach the classical scenario of a non-interacting impurity, which corresponds to a sharp peak in the spectrum at $E=0$.

\subsection{Polaron with three-body correlations}\label{sec:three-body-results}

We now go beyond previous work at finite temperature and consider the three-body ansatz. This is obtained by commuting the bare impurity four times with the Hamiltonian, yielding
\begin{widetext}
\begin{align} \nn
\cre{\mathbf{c}}{\0} & = \alpha_0 \cre c\0 
+ \gamma_0 \cre d\0 
+ \sum_\q \gamma^\q \cre{d}{\q} \ann b\q 
+ \sum_\q \alpha^\q \cre{c}{\q} \ann b\q
+ \sum_\k \alpha_\k \cre{c}{-\k} \cre b\k 
+ \sum_{\q,\k} \alpha^\q_\k \cre{c}{\q-\k} \cre b\k \ann b\q
 \\ \nn 
& + \sum_\k \gamma_\k \cre{d}{-\k} \cre b\k
+ \sum_{\q,\k} \gamma^\q_\k \cre{d}{\q-\k} \cre b\k \ann b\q
+ \frac{1}{2}\!\sum_{\q_1,\q_2} \gamma^{\q_1,\q_2} \cre{d}{\q_1+\q_2} \ann b{\q_1} \ann b{\q_2} 
+ \frac{1}{2}\!\sum_{\q_1,\q_2,\k} \gamma^{\q_1,\q_2}_\k \cre{d}{\q_1+\q_2-\k} \cre b\k \ann b{\q_1} \ann b{\q_2}
 \\ \nn 
& + \frac{1}{2}\!\sum_{\k_1,\k_2} \alpha_{\k_1,\k_2} \cre{c}{-\k_1-\k_2} \cre b{\k_1} \cre b{\k_2}
+ \frac{1}{2}\!\sum_{\q,\k_1,\k_2} \alpha^{\q}_{\k_1,\k_2} \cre{c}{\q-\k_1-\k_2} \cre b{\k_1} \cre b{\k_2} \ann b\q
+ \frac{1}{2}\!\sum_{\q_1,\q_2} \alpha^{\q_1,\q_2} \cre{c}{\q_1+\q_2} \ann b{\q_1} \ann b{\q_2}
 \\ \label{eqn:ansatz-three-body}
& + \frac{1}{2}\!\sum_{\q_1,\q_2,\k} \alpha^{\q_1,\q_2}_{\k} \cre{c}{\q_1+\q_2-\k} \cre b\k \ann b{\q_1}\ann b{\q_2}
+ \frac{1}{4}\!\sum_{\q_1,\q_2,\k_1,\k_2} \alpha^{\q_1,\q_2}_{\k_1,\k_2} \cre{c}{\q_1+\q_2-\k_1-\k_2} \cre b{\k_1} \cre b{\k_2} \ann b{\q_1} \ann b{\q_2} .
\end{align}
\end{widetext}
Here we have again taken the ideal-gas limit $\ab \to 0$ before generating the approximate variational operator.  Equation~\eqref{eqn:ansatz-three-body} contains all possible two- and three-body correlations at arbitrary temperature. The first line corresponds to the two-body ansatz \eqref{eqn:ansatz-two-body}, while the remaining lines involve terms with at least two particle or hole excitations of the medium, which are denoted by subscripts or superscripts, respectively, on the coefficients.  This ansatz represents the minimal set of terms that can describe Efimov physics across the full range of temperatures. At zero temperature, it reduces to the ansatz in Eq.~\eqref{eq:variation0} with $\ab =0$, while in the high-temperature limit, it becomes equivalent to the third-order term in the virial expansion considered in Ref.~\cite{Sun2017}.  Note that, since the particle and hole excitations correspond to identical bosons, we require the coefficients to be symmetric with respect to their permutations, e.g., $\alpha^{\q_1,\q_2}_{\k_1,\k_2} = \alpha^{\q_1,\q_2}_{\k_2,\k_1} = \alpha^{\q_2,\q_1}_{\k_1,\k_2}$, and similarly for other terms with multiple particle or hole excitations.

The resulting set of linear equations for the three-body ansatz is presented in Eq.~\eqref{eqn:three-body-equations} of Appendix~\ref{sec:eqns}.  Since the equations involve functions of multiple momentum vectors, the corresponding matrix in our numerical calculation quickly grows with the number of grid points.  Therefore, we reduce the scale of our computation by expanding the equations in spherical harmonics and then keeping only the lowest order $s$-wave term, which amounts to removing all the angular dependence of the momenta (see Appendix \ref{sec:s-wave}).  This approximation should be reasonable at low temperatures $T<T_c$, since the hole momenta are close to zero in this regime and the spectrum of Efimov trimers in Fig.~\ref{fig:EfimovSketch} only weakly depends on higher partial waves~\cite{Yoshida2018,Braaten2006}.  We have also tested the accuracy of the $s$-wave approximation for the equal-mass case using the simpler interacting-gas ansatz from Sec.~\ref{sec:interacting-results}. As shown in Appendix \ref{sec:s-wave}, we find that the approximation is accurate for weak boson-impurity interactions $1/n^{1/3}a \lesssim -2$ and produces only quantitative changes to the attractive polaron for stronger interactions while preserving the qualitative features of the spectral function.

\begin{figure}
    \centering
    \includegraphics{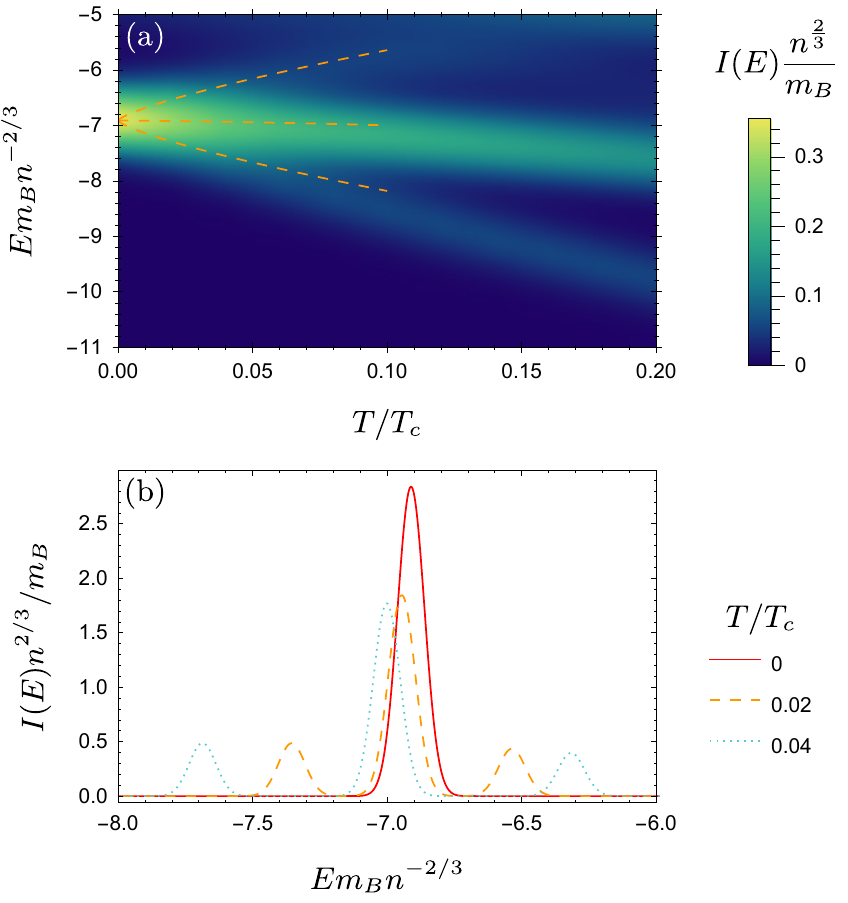}
    \caption{(a) Impurity spectral function at unitarity produced by the three-body ansatz \eqref{eqn:ansatz-three-body} with $m=m_B$, $n^{1/3}R^*=0.02$, and $\sigma=0.4 n^{2/3}/m_B$.  Only the energy range around the attractive polaron is plotted. The orange dashed lines are the predicted energies \eqref{eq:DE3body} from a low-temperature analysis.
    (b) Slices through the spectrum (a) at several low temperatures, with a narrower broadening of $\sigma=0.05 n^{2/3}/m_B$.}
    \label{fig:three-body-temperature}
\end{figure}

\begin{figure}
    \centering
    \includegraphics{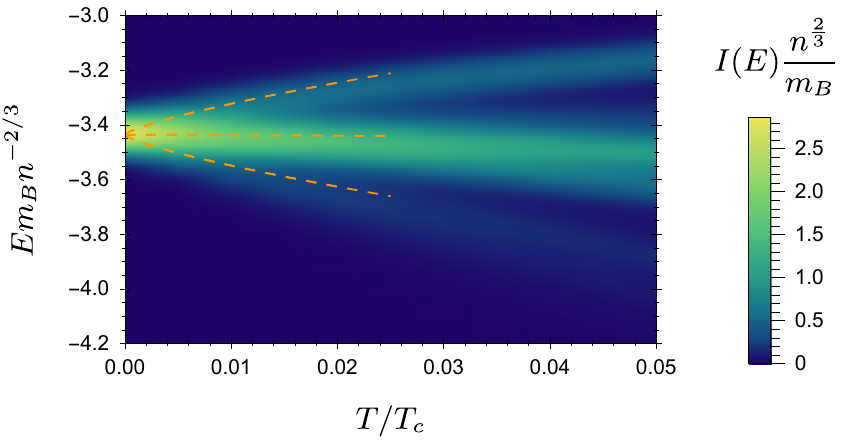}
    \caption{
    Impurity spectral function at unitarity produced by the three-body ansatz \eqref{eqn:ansatz-three-body} with $m/m_B\to\infty$, $n^{1/3}R^*=0.02$, and $\sigma=0.05 n^{2/3}/m_B$.
    The orange dashed lines are the predicted energies \eqref{eq:DE3body} from a low-temperature analysis.}
    \label{fig:three-body-static}
\end{figure}

To investigate the splitting of the attractive polaron at low temperatures, we calculate the spectral function at unitarity within an energy interval around the attractive branch, as shown in Fig.~\ref{fig:three-body-temperature}.  Crucially, we see that the ground-state polaron peak splits into \textit{three} branches as the temperature is increased from zero. Moreover, Fig.~\ref{fig:three-body-temperature}(b) shows that the ratio of spectral weights between the three peaks appears to be 1:4:1 close to zero tempeature. This is in contrast to the spectrum of the two-body ansatz in Fig.~\ref{fig:two-body-temperature}, where the attractive polaron splits into a doublet with a 1:1 ratio at low temperatures.  However, we see in both cases that a significant portion of the spectral weight of the attractive polaron shifts downwards with increasing temperature.

We can furthermore demonstrate that the behavior in Fig.~\ref{fig:three-body-temperature} is not exclusive to the equal-mass case.  In Fig.~\ref{fig:three-body-static}, we plot the low-temperature spectral function for an infinitely heavy impurity at unitarity, and we once again observe a splitting of the attractive polaron into three branches.  The fixed-impurity case is particularly instructive since there is no angular dependence of momenta in impurity-boson scattering due to the lack of recoil, and thus our $s$-wave approximation for the three-body ansatz becomes exact. In addition, there are no Efimov trimer bound states when the impurity mass is infinite~\cite{Efimov1973}.  We thus conclude that the observed triple splitting is independent of mass ratio and Efimov physics.

\section{Origin of polaron splitting}\label{sec:origin}

Our results for the finite-temperature spectral function show that the behavior of the attractive branch is dependent on the choice of variational ansatz, thus raising serious doubts about whether the splitting of the attractive branch is physical.  Therefore, in this section, we aim to gain further insight into this splitting by determining how it arises in the low-temperature limit, $T\ll T_c$ and $T \ll |E_{\rm att}|$, where $E_{\rm att}$ is the attractive polaron energy at zero temperature.

\begin{figure}
    \centering
    \includegraphics[width=\linewidth]{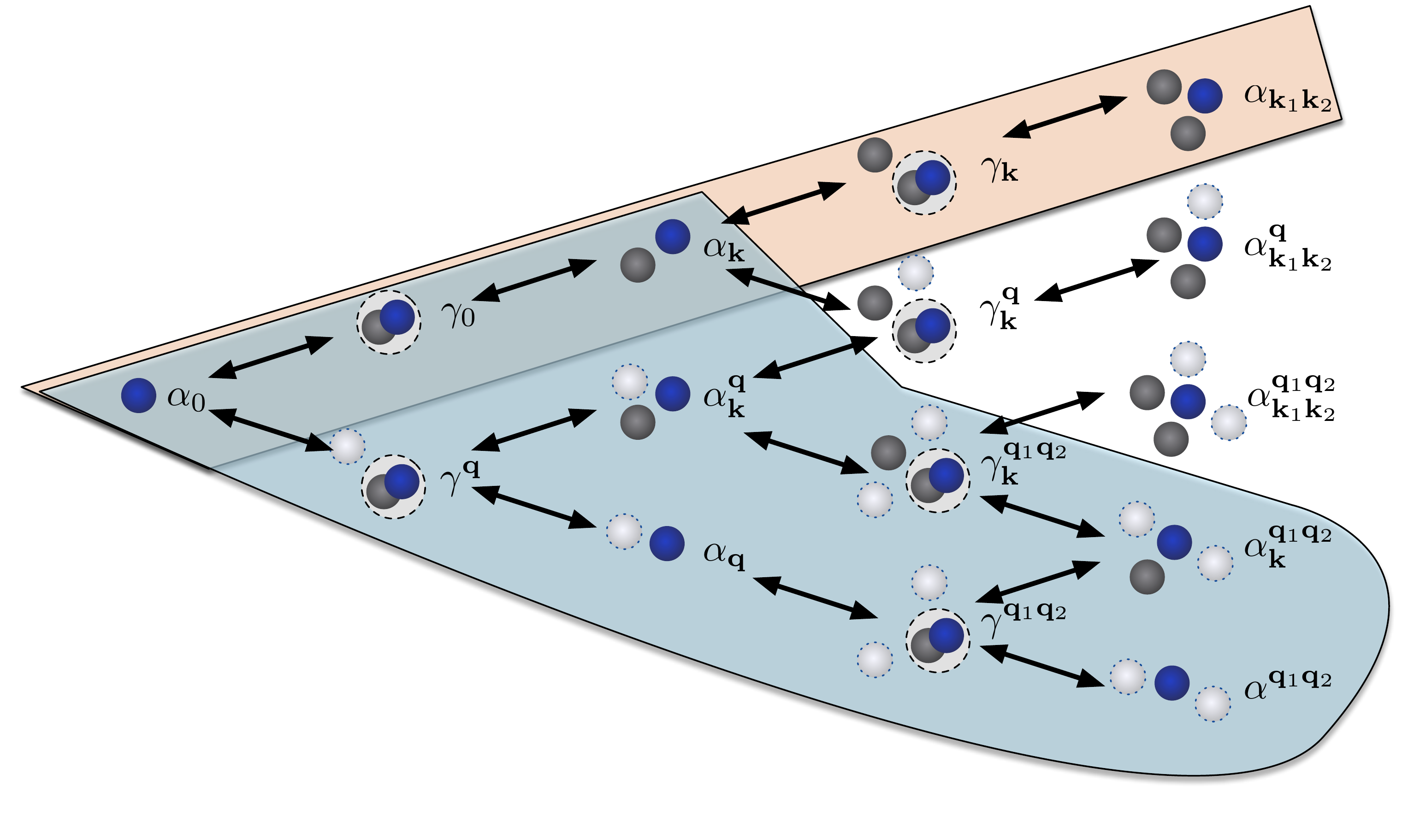}
    \caption{Connections between the various coefficients in the three-body ansatz \eqref{eqn:ansatz-three-body}. The various terms are illustrated by the presence of the impurity (blue circle), particles (dark gray circles), holes (light gray circles), and the closed-channel dimer (dark gray and blue circles). The blue shaded area contains those terms responsible for the splitting into three branches, while the orange area are those terms responsible for Efimov physics at zero temperature.}
    \label{fig:coeffs}
\end{figure}

To this end, we consider the structure of the variational equations and how the different terms in the variational ansatz are coupled to one another by the Hamiltonian, as shown in Fig.~\ref{fig:coeffs}. While the number of coefficients grows rapidly as we include higher order correlations, we can identify two important classes: the zero-temperature terms that contain Efimov physics, and the minimal set of finite-temperature terms that produce splitting of the attractive branch. In particular, we find that the structure of the splitting is captured by 1-particle terms with different numbers of hole excitations (blue shaded area of Fig.~\ref{fig:coeffs}).
Additional details can be found in Appendix~\ref{sec:spec}.

\begin{figure}[t]
    \raggedright
    \import{./}{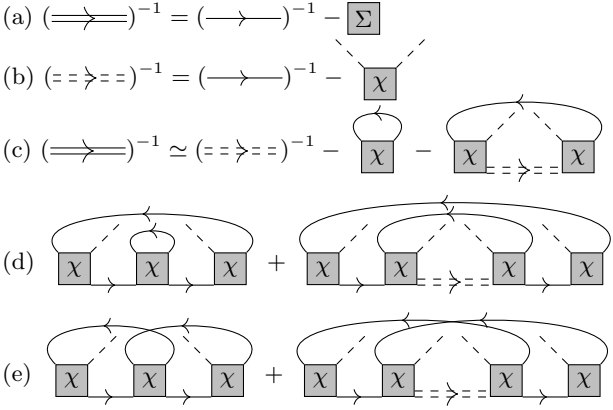}
    \caption{Diagrammatic approach to quasiparticle branch splitting. (a) Dyson equation for the impurity Green's function (double line) in terms of the bare impurity Green's function (single line) and the self energy (shaded box). (b) Dyson equation at $T=0$. (c) Dyson equation at $0<T\ll T_c$, where we only include the most important diagrams for the splitting with one hole excitation. 
    (d) Lowest order diagrams where we dress the bare impurity propagator inside the $T=0$ impurity propagator (double dashed line) of panel (c). Diagrams where hole propagators cross each other, as depicted in (e), also appear at the same order.}
    \label{fig:diagrams}
\end{figure}

To illustrate this point, we now present an alternative diagrammatic argument (Fig.~\ref{fig:diagrams}), where we again consider the impurity operator at rest for simplicity.  The retarded impurity Green's function (or propagator) $G(E)$ satisfies the Dyson equation illustrated in Fig.~\ref{fig:diagrams}(a):
\begin{align}
    G^{-1}(E)=G_0^{-1}(E)-\Sigma(E)=E-\Sigma(E) .
\end{align}
Here, the bare impurity Green's function $G_0(E)=1/E$, while the self energy $\Sigma(E)$ contains the effect of the impurity-boson interactions.  The quasiparticle branches then correspond to solutions of ${\rm Re}[G^{-1}(E)]=0$ since the impurity spectral function is related to $G$ via
\begin{align}
    A(E)=-{\rm Im}\,[G(E+i0)]/\pi,
\end{align}
where the quantity $+i0$ slightly shifts the poles of the Green's function in the complex plane. This representation is equivalent to the spectral function in Eq.~\eqref{eq:Aomega}.

At zero temperature, we can write the Dyson equation as
\begin{align}
    G^{-1}(E)=E-n_0\chi(E),
\end{align}
as illustrated in panel (b) of Fig.~\ref{fig:diagrams}.  Here, the self energy $\Sigma(E)=n_0\chi(E)$ is the sum of diagrams where the impurity excites at least one boson out of the condensate. Note that since we are considering the limit $\ab \to 0$, there is no quantum depletion of the condensate, and thus the self energy has to begin and end with condensate lines.

At finite temperature, the impurity can also interact with thermally excited bosons, which gives rise to a new class of diagrams. These either begin, end, or both begin and end with thermal (hole) excitations. However, as illustrated in Fig.~\ref{fig:diagrams}(c), the diagrams that begin \textit{and} end with thermally excited bosons involve the $T=0$ propagator from Fig.~\ref{fig:diagrams}(b). Importantly, close to the pole of the $T=0$ Green's function, these diagrams are divergent and hence these form the most important corrections to the self energy at low temperature. In Fig.~\ref{fig:diagrams}(c) we consider the correction to the self energy at lowest order in the density of thermally excited bosons, $\nx$, where we note that at low temperature we can neglect the hole momentum. Combining the diagrams in Fig.~\ref{fig:diagrams}(b) and (c) we therefore have
\begin{align}
    G^{-1}(E)\simeq E-n_0\chi(E)-\frac{\nx}{\chi^{-1}(E)-\frac{n_0}E}.
    \label{eq:Gone}
\end{align}
When $\chi(E)$ corresponds to the T matrix, the condition $G^{-1}(E) = 0$ is equivalent to Eq.~\eqref{eq:extTmat} in the limit where $f_\q \to 0$ such that we can set the hole momentum to zero in the kinetic terms and we are left with $\nx = \sum_\q f_\q$. An equation similar to Eq.~\eqref{eq:Gone} was also analyzed diagrammatically in Ref.~\cite{Guenther2018}.

We are now in a position to solve for the quasiparticle energy at low $T$, assuming a single hole excitation (like in the two-body ansatz).  Expanding Eq.~\eqref{eq:Gone} around the $T=0$ pole, we find the two solutions
\begin{align}
E\simeq E_{\rm att}\left[1\pm Z_{\rm att} (n_{\rm ex}/n_0)^{1/2}\right].
\label{eq:DE2body}
\end{align}
In the case of an ideal Bose gas at sufficiently low temperatures, we have $n_{\rm ex}/n_0=(T/T_c)^{3/2}$, while it scales as $T^2$ when $\ab$ is appreciable and the quantum depletion exceeds the thermal depletion~\cite{Shi1998}. In Eq.~\eqref{eq:DE2body}
\begin{align}
Z_{\rm att}=\left[1-\left.\frac{\partial {\rm Re}[\Sigma]}{\partial
  E}\right|_{E=E_{\rm att};T=0}\right]^{-1}
\end{align}
is the zero-temperature quasiparticle residue~\cite{Fetter}, i.e., it is the squared overlap of the polaron ground state with the non-interacting ground state.  The two solutions in Eq.~\eqref{eq:DE2body} are illustrated in Fig.~\ref{fig:two-body-temperature}, and it is seen that they fit very well with the numerical results within the two-body ansatz. Our expression for the Green's function in Eq.~\eqref{eq:Gone} also allows us to see that the two branches have equal residues at low temperature, far below $T_c$.

One might assume that it is sufficient to consider diagrams involving up to one hole excitation when determining the low-temperature behavior of the Bose polaron.  However, close to the $T=0$ pole of the Green's function, it is important to note that terms involving several hole excitations can add up to yield contributions to the polaron energy at the same order in $\nx$ as that found in Eq.~\eqref{eq:DE2body}.  The key observation is that while $\nx\to0$ as $T\to0$, the denominator in Eq.~\eqref{eq:Gone} also vanishes at the pole, and taking the limit is therefore non-trivial. To see how this works, consider now the diagrams where we dress the bare propagator inside the $T=0$ diagrams (double-dashed propagator) in the last term of Fig.~\ref{fig:diagrams}(c) with thermal excitations.  This amounts to replacing $\frac 1E\longrightarrow\frac1{E-\frac{\nx}{\chi^{-1}(E)-n_0/E}}$ in the last term of the denominator of Eq.~\eqref{eq:Gone}, where the first non-trivial such diagram is shown in Fig.~\ref{fig:diagrams}(d). However, along with each such diagram, we have additionally a ``crossed'' diagram such as that shown in Fig.~\ref{fig:diagrams}(e). Therefore, in total we replace
\begin{align}
    \frac 1E\longrightarrow \frac1{E-\frac{2\nx}{\chi^{-1}(E)-n_0/E}}.
\end{align}
This finally yields the expression
\begin{align}
    G^{-1}(E)\simeq
  E-n_0\chi(E)-\frac{\nx}{\chi^{-1}(E)-\frac{n_0}{E-2\frac{\nx}{\chi^{-1}(E)-n_0/E}}},
\label{eq:Gtwo}
\end{align}
which now includes up to two simultaneous thermally excited bosons. This is equivalent to a variational ansatz with up to two hole excitations, as depicted in Fig.~\ref{fig:coeffs}.

An analysis of Eq.~\eqref{eq:Gtwo} yields three solutions for the attractive polaron energy at low temperature.  Two of these are symmetric around the atractive polaron,
\begin{align}
E\simeq E_{\rm att}\left[1\pm \sqrt{3Z_{\rm att}} (n_{\rm ex}/n_0)^{1/2}\right],
\label{eq:DE3body}
\end{align}
while the third solution remains $E\simeq E_{\rm att}$ at order $(n_{\rm ex}/n_0)^{1/2}$.  These solutions are illustrated in Figs.~\ref{fig:three-body-temperature}(a) and \ref{fig:three-body-static}, and again we find a good agreement with our numerical results.  Note that simply expanding Eq.~\eqref{eq:Gtwo} yields the residue instead of its square root in Eq.~\eqref{eq:DE3body}. However, numerically we find a near perfect agreement with Eq.~\eqref{eq:DE3body} at sufficiently low temperatures, and we attribute this difference to processes beyond those taken into account in Eq.~\eqref{eq:Gtwo} (see Appendix~\ref{sec:spec}).  These additional processes do not change the scaling with $\nx$ nor the structure of the splitting. In both cases, the residues of the three branches are found to have the relationship 1:4:1, which is seen to well match the result in Fig.~\ref{fig:three-body-temperature}(b).

The procedure described above can be extended to arbitrary order by including additional holes in the ansatz operator, effectively leading to the possibility of replacing the bare impurity propagator in the last denominator by that dressed by thermal excitations. Carrying out such a procedure, one arrives at a pole condition in the form of a continued fraction of ever increasing order:
\begin{align} \nn
  E=n_0\chi(E)+\frac{\nx}{\chi^{-1}(E)-\frac{n_0}{E-2\frac{\nx}{\chi^{-1}(E)-\frac{n_0}{E-3\frac{\nx}{\chi^{-1}(E)-\frac{n_0}{E-4\cdots}}}}}}.
\end{align}
Note that, due to the presence of crossed diagrams such as in Fig.~\ref{fig:diagrams}(e), this does not simply correspond to a self-consistent expansion of the impurity propagator.

From the above considerations, we conclude that the splitting of the attractive polaron branch depends on the number of hole excitations included in the variational ansatz.  Importantly, the structure of the equations allows us to make some general observations on the nature of the attractive polaron branch at finite temperature. First of all, by continuing the process described above, we can immediately see that the number of attractive branches is simply one larger than the number of holes in the ansatz. Second, quite remarkably we see that the structure of the equations is such that the quasiparticle branches only depend on $E_{\rm att}$ and $Z_{\rm att}$, and not on the details of the interactions and mass ratio. Finally, the correction to the energy scales as $|E_{\rm att}|\sqrt{n_{\rm ex}/n_0}$ for all branches, and we therefore expect that the exact polaron forms a single broad peak at finite temperature, with a peak width given by
\begin{align} \label{eq:gamma}
\Gamma\propto |E_{\rm att}|\sqrt{n_{\rm ex}/n_0}.
\end{align}
Thus, $\Gamma$ scales as $T^{3/4}$ in the ideal gas limit. Indeed, we expect such a scaling to hold even for finite $\ab$, provided we are in the regime where $g_B n \ll T \ll |E_{\rm att}|$, $T_c$.  On the other hand, in the opposite regime $T \ll g_B n$, Eq.~\eqref{eq:gamma} implies that $\Gamma$ scales linearly with $T$ since $\nx \propto T^2$ at low temperatures, according to Popov theory~\cite{Shi1998}.

We can understand the splitting as arising from the two-fluid nature of the Bose gas at low temperatures. Here, a necessary ingredient is that particles from one fluid can be converted into the other, as we can see from Fig.~\ref{fig:diagrams}.  Moreover, we require one of the fluids to have an occupation of modes at low but finite momentum.  For instance, if we replaced the hole lines in Fig.~\ref{fig:diagrams} by condensate lines of a second BEC (formed from particles of the same mass and intraspecies interaction strength), then the correction to the $T=0$ self energy would not be one particle irreducible. Thus, the pole condition for the impurity Green's function would not involve an infinite continued fraction.

\section{Effect of boson interactions}\label{sec:interacting-results}

Throughout this paper, we have taken the ideal-gas limit when constructing and applying the variational ansatzes. We now explicitly investigate the effect of weak interactions in the Bose medium.  These are not expected to strongly affect the ground-state energy as long as another short-distance scale, such as $R^*$, cuts off the three-body spectrum~\cite{Yoshida2018}, but they can impact the low-energy physics as well as the variational ansatz we consider.

We can construct an ansatz operator for the interacting Bose gas in the same manner as above. However, since the bare boson creation operator involves both creation and annihilation Bogoliubov operators, it has a slightly different form. By commuting the bare impurity with the Hamiltonian twice, we thus produce an ansatz containing all combinations of two Bogoliubov operators,
\begin{align} \nn
\cre{\mathbf{c}}{\0} ={}& \alpha_0 \cre c\0 + \gamma_0 \cre d\0 + \sum_\q \gamma^\q \cre{d}{\q} \ann\beta \q + \sum_\k \gamma_\k \cre{d}{-\k} \cre \beta \k\\ \nn
&+ \sum_\q \alpha^\q \cre{c}{\q} \ann\beta \q + \sum_\k \alpha_\k \cre{c}{-\k} \cre\beta \k + \sum_{\q,\k} \alpha^\q_\k \cre{c}{\q-\k} \cre{\beta}{\k} \ann{\beta}{\q}\\ \nn
&+ \frac{1}{2} \sum_{\q_1,\q_2} \alpha^{\q_1,\q_2} \cre{c}{\q_1+\q_2} \ann\beta{\q_1} \ann\beta{\q_2}\\ \label{eqn:ansatz-bog}
&+ \frac{1}{2} \sum_{\k_1,\k_2} \alpha_{\k_1,\k_2} \cre{c}{-\k_1-\k_2} \cre\beta{\k_1} \cre\beta{\k_2}.
\end{align}
In the limit of zero temperature, this interacting-gas ansatz corresponds to Eq.~\eqref{eq:variation0}, which is the minimal ansatz that correctly reproduces weak-coupling perturbation theory~\cite{Novikov2009,Christensen2015} beyond mean-field for the interacting Bose gas.  On the other hand, above the critical temperature, this ansatz simply reduces to the two-body ansatz.  As such, this allows us to investigate correlations of the Bose polaron that are of intermediate complexity between the two-body and three-body ansatzes.  The system of equations resulting from the minimization procedure may be found in Appendix~\ref{sec:eqns}.

\begin{figure}
    \centering
    \includegraphics{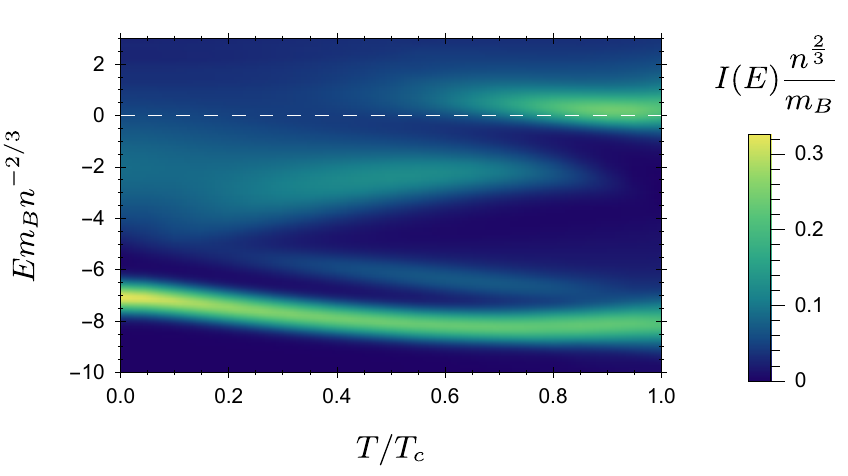}
    \caption{Impurity spectral function produced by the ansatz \eqref{eqn:ansatz-bog} plotted with respect to temperature at $1/n^{1/3}a=0$. Here we take $n^{1/3}\ab=0.003$, $n^{1/3}R^*=0.02$, $m=m_B$ and $\sigma=0.4 n^{2/3}/m_B$, 
      which correspond to the parameters of the Aarhus experiment \cite{Jorgensen2016}.}
    \label{fig:bog-T}
\end{figure}

\begin{figure*}
    \centering
    \includegraphics{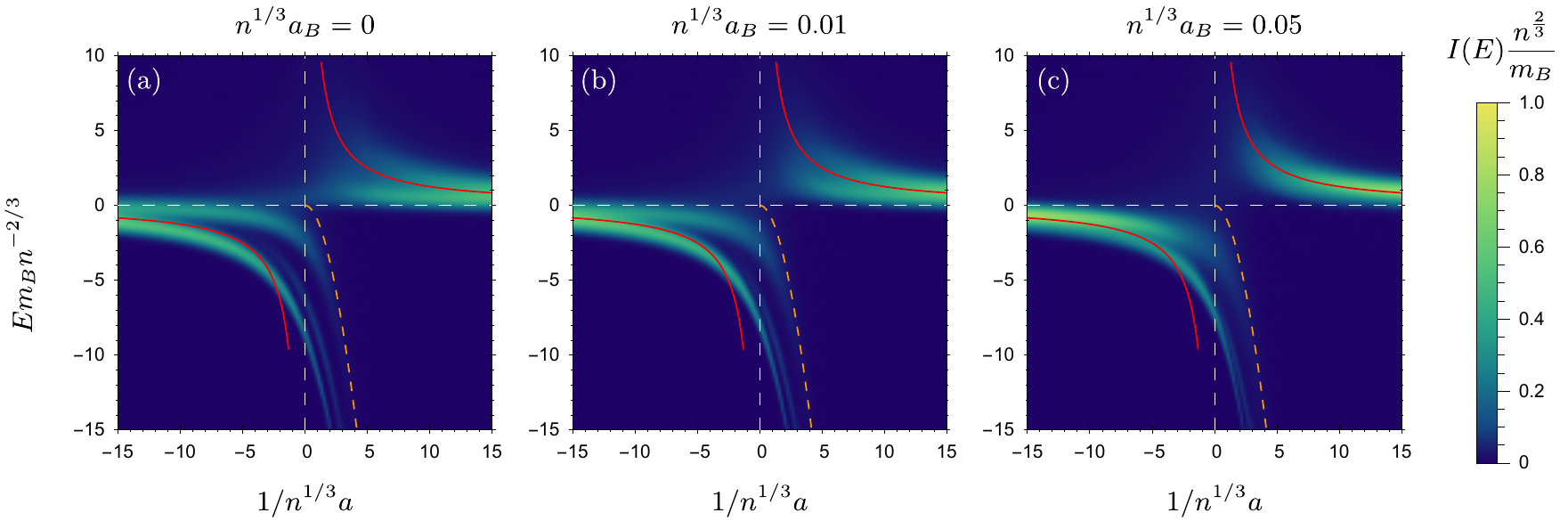}
    \caption{Impurity spectra produced by the ansatz \eqref{eqn:ansatz-bog} at $T=0.5T_c$ with boson-boson scattering length of (a) $n^{1/3}\ab=0$, (b) $n^{1/3}\ab=0.01$, and (c) $n^{1/3}\ab=0.05$. The solid red lines are the mean-field energy \eqref{eqn:meanfield}, while the dashed orange lines are the dimer energy $-\eb$. Here we take $n^{1/3}R^*=0.02$, $m=m_B$ and $\sigma=0.4 n^{2/3}/m_B$.}
    \label{fig:bog-ainv}    
\end{figure*}

The zero-temperature spectrum of the finite-$a_B$ ansatz features both attractive and repulsive polaron branches and thus qualitatively resembles the spectrum obtained from the two-body ansatz in Fig.~\ref{fig:two-body}(a). The main difference is the existence of a many-body continuum of states between the polaron branches, which requires at least three-body correlations in order to be captured~\cite{Jorgensen2016}. This continuum manifests as a broad feature with small spectral weight even in the strong-coupling regime, as shown in Fig.~\ref{fig:bog-T}.  Note that such a continuum is also present in the spectrum for the three-body ansatz, but at energies higher than that plotted in Fig.~\ref{fig:three-body-temperature}.

Increasing the temperature from zero, we find that the attractive polaron remains as a single peak rather than splitting into several branches, unlike what we found previously.  This is because the interacting-gas ansatz \eqref{eqn:ansatz-bog} only includes up to two-body correlations between the impurity and the thermal cloud, while containing up to three-body correlations between the impurity and the condensate. Thus, the impurity effectively interacts differently with the two fluids such that the high-order pole structure from Sec.~\ref{sec:origin} no longer applies.  However, we instead find that the continuum changes its structure at low temperature and splits into two peaks, as shown in Fig.~\ref{fig:bog-T}. Thus, the polaron splitting is transferred to the excited states in the spectrum, where the dressed impurity has shed a Bogoliubov excitation and now only contains up to two-body correlations.  Such behavior is once again observed for a range of different boson-impurity interactions (Fig.~\ref{fig:bog-ainv}) and mass ratios.

In addition to confirming that the character of the splitting depends on the approximation used, our results show that the arguments of Sec.~\ref{sec:origin} can apply to a continuum of states, not just a single well-defined quasiparticle.  Indeed, we find that the splitting in Fig.~\ref{fig:bog-T} can be approximately fit with the low-temperature expression \eqref{eq:DE2body} using reasonable values for the energy and residue of the continuum.  Thus, we expect our conclusions regarding the attractive polaron to hold even when it consists of multiple eigenstates at zero temperature. Such a scenario is likely to occur in the ideal gas limit, since the ground-state residue is predicted to vanish~\cite{Christensen2015,Yoshida2018} as we include more and more excitations of the condensate. At the same time, the gap between the ground and excited states will tend to zero, such that the spectrum of the interacting-gas ansatz converges to that of the previous ansatzes.

As the temperature approaches $T_c$, we see in Fig.~\ref{fig:bog-T} that the continuum of states vanishes since only two-body correlations are included in the thermal gas above $T_c$.  This behavior could be an artifact of the approximations in the interacting-gas ansatz and thus further investigations are required to determine how the spectrum changes around $T_c$.  However, it appears unlikely that the attractive polaron has an energy minimum exactly at $T_c$, as predicted in Ref.~\cite{Guenther2018}, since we see that the position of this minimum depends on the approximation we consider.

Figure~\ref{fig:bog-ainv} shows how the finite-temperature spectrum changes as we increase $n^{1/3}\ab$. Generically we find that the attractive branch is shifted upwards, similarly to what was observed for the ground state at zero temperature~\cite{Levinsen2015,Yoshida2018}. We also find that the splitting at negative energies disappears at weak boson-impurity interactions for sufficiently large $n^{1/3}\ab$. Such behavior is consistent with weak-coupling perturbation theory in the regime $\ab \gg a$, where no splitting is observed~\cite{Levinsen2017}.  Finally, we see that the spectral weight around zero energy becomes suppressed with increasing $n^{1/3}\ab$. As we discussed in Sec.~\ref{sec:ansatzes}, this zero-energy peak is primarily composed of low-momentum bosons that were scattered out of the condensate by the impurity. Such scattering processes are Bose enhanced by the presence of thermal bosons and this shows up in the singular nature of the thermal distribution $f_\k$ at low momentum.  However, the Bose enhancement also depends on the low-energy density of states of the Bogoliubov excitations, and this is reduced with increasing $n^{1/3}\ab$, since the Bogoliubov dispersion changes from quadratic to linear at low momentum. Thus, the phase space for the scattering processes is also reduced at low momentum, and this decreases the spectral weight around zero energy.

\section{Conclusion}\label{sec:conclusion}

In this paper, we have investigated the Bose polaron using a recently developed finite-temperature variational approach to impurity dynamics~\cite{Liu2019}. At zero temperature, we have compared our results for the polaron ground-state energy with recent quantum Monte Carlo calculations~\cite{Pena-Ardila2019} and with the re-analysed data~\cite{Pena-Ardila2019} from the Aarhus experiment~\cite{Jorgensen2016}.  Crucially, in both cases, the agreement was excellent without the need for any adjustable parameters once we had carefully matched the length scale associated with Efimov physics.  Our results therefore further support the finding that the ground-state energy of the Bose polaron in the strongly interacting regime is a universal function of the length scale associated with Efimov physics~\cite{Yoshida2018}.  The validity of our variational approach can be further tested by performing a detailed comparison with recent experimental results for the spin dynamics of localized Bose polarons~\cite{Schmidt2018}.

Our findings complement recent theory predictions for localized impurities in a Bose gas, where it was found that the impurity induces a strong boson-boson repulsion even in the absence of any direct repulsive interaction~\cite{Shi2018Multibody,Yoshida2018PRA}. Therefore, a picture is emerging where the ground state of the Bose polaron with short range interactions appears well described by calculations including only a few excitations of the medium. This is qualitatively very different from impurities with long-range interactions such as Rydberg polarons~\cite{Camargo2018}, which form ``super-polaronic'' states dressed by arbitrarily many excitations.

At finite temperature, we focused on the splitting of the attractive polaron quasiparticle that has been predicted to occur~\cite{Guenther2018} below the critical temperature for Bose-Einstein condensation. We demonstrated that, generically, the number of attractive branches at finite temperature is simply set by the number of hole excitations of the thermal cloud, and we illustrated this using variational ansatzes with one or two hole excitations.  Furthermore, we explained the origin of the splitting as being due to the two-fluid nature of the medium --- the superfluid and the thermal component --- and the fact that the two fluids can exchange particles. Consequently, the impurity self energy has the form of an infinite continued fraction in the vicinity of the ground state. We emphasize that this is a highly non-perturbative result which applies even in the weak coupling limit, as long as $|a|\gtrsim\ab$.  Interestingly, the structure of the continued fraction implies that the energies of the split branches solely depend on the impurity-boson interaction strength and mass ratio through the energy and residue of the $T=0$ polaron.

Our results suggest that the attractive polaron quasiparticle in the exact spectrum will consist of a single thermally broadened peak. Furthermore, using a diagrammatic analysis, we have argued that the width of this peak scales as $\sqrt{n_{\rm ex}/n_0}$ at small $T$. Hence, the width is expected to scale as $T^{3/4}$ in the limit where $\ab$ can be neglected, while it should scale linearly in $T$ when the quantum depletion exceeds the thermal depletion.

As illustrated in this work, the variational approach introduced in Ref.~\cite{Liu2019} represents a clear and straightforward manner in which one can systematically go beyond ladder type approximations to the polaron problem. These ideas can potentially be applied to other systems featuring quasiparticles dressed by (gapless) thermal excitations of a medium.  In particular, our diagrammatic arguments relied solely on the fact that the impurity interacts with two fluids that can exchange particles, and therefore we expect them to hold for an impurity interacting with any medium that features a broken symmetry phase.

As an intriguing outlook, in the case of an electron in a quark-gluon plasma~\cite{Klimov1982} a one-loop calculation~\cite{Baym1992} has demonstrated that the bare electron splits into two branches at finite temperature. Therefore, one is naturally led to speculate that a higher-order variational calculation might yield multiple splittings, and that an exact calculation would yield a broadened peak rather than split quasiparticle branches.

\acknowledgments 
We are grateful to P.~Massignan for fruitful discussions, and we thank L.~A.~Pe\~{n}a Ardila for providing us with the QMC and experimental data from Ref.~\cite{Pena-Ardila2019}. JL is supported through the Australian Research Council Future Fellowship FT160100244. BF is supported through an Australian Government Research Training Program (RTP) Scholarship. We also acknowledge support from the Australian Research Council Centre of Excellence in Future Low-Energy Electronics Technologies (CE170100039).

\appendix

\begin{widetext}

\section{Numerical methods}\label{sec:numerics}

The systems of equations \eqref{eqn:two-body-equations}, \eqref{eqn:three-body-equations} and \eqref{eqn:bog-equations} are all linear integral equations. We convert the sums to integrals in spherical coordinates, making use of spherical symmetry to integrate out unneeded variables. We use Gauss-Legendre quadrature for the integrals. For the momentum integrals, the Gauss-Legendre abscissas were transformed as $\frac{2\alpha}{x+1}-\alpha$ to go from a finite to a long-tailed domain, where $\alpha$ is a scaling factor which can be chosen to promote convergence~\cite{Numerical_Recipes_Gaussian_Quadrature}.

The number of abscissas for hole momentum, particle momentum, and angle, as well as the ultraviolet cutoff $\Lambda$, were chosen to ensure convergence of the broadened spectral function. For a given set of physical parameters, this was done by increasing these integration parameters until the broadened spectral function did not visibly change in the energy region of interest.

We performed an additional step of preprocessing on the linear integral equations before we evaluated their eigenvalues and eigenvectors. This preprocessing step simultaneously ensured proper normalisation of the eigenvectors, and symmetrized the matrix which made evaluating the eigensystem faster.
Our normalisation condition is $\langle \ann{\mathbf c}\0 \cre{\mathbf c}\0\rangle$, which is not simply the amplitude squared of the coefficients $\alpha$ and $\gamma$. Explicitly, the normalization condition for the two-body ansatz \eqref{eqn:ansatz-two-body} is
\begin{equation}
\abs{\alpha_0}^2 
+ \abs{\gamma_0}^2 
+ \sum_\q f_\q \abs{\gamma^\q}^2
+ \sum_\k (1+f_\k) \abs{\alpha_\k}^2 
+ \sum_\q f_\q \abs{\alpha^\q}^2
+ \sum_{\q,\k} f_\q (1+f_\k) \abs{\alpha^\q_\k}^2 
= 1.
\label{eq:norm}
\end{equation}
The normalization conditions for the other ansatzes are in Appendix \ref{sec:eqns}.

Consider then the variational coefficients $\alpha$ and $\gamma$ as forming a vector $\vect{v}$. We can define a diagonal matrix $\mathbf{D}$ consisting of the terms from the normalisation condition such that $\vect{v}^\dagger\mathbf{D}\vect{v}$ is precisely the left hand side of Eq.~\eqref{eq:norm}. Let the as-presented linear equations be $E\vect{v}=\mathbf{K}\vect{v}$. This can be transformed to $E\mathbf{D}^{1/2}\vect{v} = (\mathbf{D}^{1/2}\mathbf{K}\mathbf{D}^{-1/2})\mathbf{D}^{1/2}\vect{v}$. We solved the now-symmetric matrix $(\mathbf{D}^{1/2}\mathbf{K}\mathbf{D}^{-1/2})$ for the normalised eigenvectors $\mathbf{D}^{1/2}\vect{v}$ which have the same eigenvalues $E$ as the original system, as described in Ref.~\cite{Numerical_Recipes_Integral_Equations}.

\section{Variational equations}\label{sec:eqns}

\subsection{Three-body ansatz}\label{sec:three-body-eqns}

In the three-body ansatz, we use the following impurity creation operator
\begin{equation}
\begin{aligned}
\cre{\mathbf{c}}{\0} &= \alpha_0 \cre c\0 
+ \gamma_0 \cre d\0 
+ \sum_\q \gamma^\q \cre{d}{\q} \ann b\q 
+ \sum_\q \alpha^\q \cre{c}{\q} \ann b\q
+ \sum_\k \alpha_\k \cre{c}{-\k} \cre b\k 
+ \sum_{\q,\k} \alpha^\q_\k \cre{c}{\q-\k} \cre b\k \ann b\q
+ \sum_\k \gamma_\k \cre{d}{-\k} \cre b\k \\
& + \sum_{\q,\k} \gamma^\q_\k \cre{d}{\q-\k} \cre b\k \ann b\q
+ \frac{1}{2}\!\sum_{\q_1,\q_2} \gamma^{\q_1,\q_2} \cre{d}{\q_1+\q_2} \ann b{\q_1} \ann b{\q_2} 
+ \frac{1}{2}\!\sum_{\q_1,\q_2,\k} \gamma^{\q_1,\q_2}_\k \cre{d}{\q_1+\q_2-\k} \cre b\k \ann b{\q_1} \ann b{\q_2}
+ \frac{1}{2}\!\sum_{\k_1,\k_2} \alpha_{\k_1,\k_2} \cre{c}{-\k_1-\k_2} \cre b{\k_1} \cre b{\k_2} \\
& + \frac{1}{2}\!\sum_{\q,\k_1,\k_2} \alpha^{\q}_{\k_1,\k_2} \cre{c}{\q-\k_1-\k_2} \cre b{\k_1} \cre b{\k_2} \ann b\q
+ \frac{1}{2}\!\sum_{\q_1,\q_2} \alpha^{\q_1,\q_2} \cre{c}{\q_1+\q_2} \ann b{\q_1} \ann b{\q_2}
+ \frac{1}{2}\!\sum_{\q_1,\q_2,\k} \alpha^{\q_1,\q_2}_{\k} \cre{c}{\q_1+\q_2-\k} \cre b\k \ann b{\q_1}\ann b{\q_2} \\
&+ \frac{1}{4}\!\sum_{\q_1,\q_2,\k_1,\k_2} \alpha^{\q_1,\q_2}_{\k_1,\k_2} \cre{c}{\q_1+\q_2-\k_1-\k_2} \cre b{\k_1} \cre b{\k_2} \ann b{\q_1} \ann b{\q_2}.
\label{eq:appa3body}
\end{aligned}
\end{equation}
Here we take the sums such that, in any given term, the medium creation and annihilation operator momenta are different and non-zero.  The corresponding system of linear equations is
\begin{subequations}
	\begin{align}
	E\alpha_0 &= g \sqrt{n_0} \gamma_0 + g\sum_{\q} f_{\q} \gamma^{\q} \\
	E\gamma_0 &= g \sqrt{n_0} \alpha_0 + \nu \gamma_0 + g \sum_{\k} (1+f_{\k}) \alpha_{\k} \\
	E\gamma^\q &= g \alpha_0 + (\epsilon_{\q,d}+\nu-\eqb) \gamma^{\q} + g \sqrt{n_0} \alpha^\q + g\sum_{\k} (1+f_{\k}) \alpha^\q_{\k} \\
	E\alpha^\q &= g\sqrt{n_0} \gamma^\q + (\eq - \eqb)\alpha^\q + \frac{g}{2} \sum_{\p} f_{\p} (\gamma^{\q,\p}+\gamma^{\p,\q}) \\
	E\alpha_\k &= g \gamma_0 + (\ek + \ekb) \alpha_\k + g  \sqrt{n_0} \gamma_\k + g\sum_{\q}f_{\q} \gamma^{\q}_\k \\
	E\alpha^\q_\k &= g \gamma^\q + (\epsilon_{\q-\k} + \ekb - \eqb) \alpha^\q_\k + g \sqrt{n_0} \gamma^\q_\k + \frac{g}{2} \sum_{\p} f_{\p} (\gamma^{\q,\p}_\k+\gamma^{\p,\q}_\k) \\
	E\gamma_\k &= g\sqrt{n_0} \alpha_\k + (\epsilon_{\k,d} +\nu + \ekb) \gamma_\k + \frac{g}{2}\sum_{\p}(1+f_{\p})(\alpha_{\k,\p}+\alpha_{\p,\k}) \\
	E\gamma^\q_\k &= g \alpha_\k + g\sqrt{n_0} \alpha^\q_\k + (\epsilon_{\q-\k,d}+\nu + \ekb-\eqb)\gamma^\q_\k + \frac{g}{2} \sum_{\p} (1+f_{\p}) (\alpha^\q_{\k,\p}+\alpha^\q_{\p,\k}) \\
	E\gamma^{\q_1,\q_2} &= g (\alpha^{\q_1}+\alpha^{\q_2}) + (\epsilon_{\q_1+\q_2,d}+\nu-\epsilon_{\q_1}^B-\epsilon_{\q_2}^B)\gamma^{\q_1,\q_2} + g \sqrt{n_0} \alpha^{\q_1,\q_2} + g \sum_{\p} (1+f_{\p}) \alpha^{\q_1,\q_2}_{\p} \\
	E\gamma^{\q_1,\q_2}_\k &= g(\alpha^{\q_1}_\k+\alpha^{\q_2}_\k) + (\epsilon_{\q_1+\q_2-\k,d}+\nu + \ekb-\epsilon_{\q_1}^B - \epsilon_{\q_2}^B) \gamma^{\q_1,\q_2}_\k + g \sqrt{n_0} \alpha^{\q_1,\q_2}_\k + \frac{g}{2}\sum_{\p} (1+f_{\p}) (\alpha^{\q_1,\q_2}_{\k,\p}+\alpha^{\q_1,\q_2}_{\p,\k}) \\
	E\alpha_{\k_1,\k_2} &= g (\gamma_{\k_1}+\gamma_{\k_2}) + (\epsilon_{\k_1+\k_2}+\epsilon_{\k_1}^B+\epsilon_{\k_2}^B)\alpha_{\k_1,\k_2} \\
	E\alpha^\q_{\k_1,\k_2} &= g (\gamma^\q_{\k_1}+\gamma^\q_{\k_2}) + (\epsilon_{\q-\k_1-\k_2}+\epsilon_{\k_1}^B+\epsilon_{\k_2}^B-\eqb)\alpha^\q_{\k_1,\k_2} \\
	E\alpha^{\q_1,\q_2} &= g\sqrt{n_0} \gamma^{\q_1,\q_2} + (\epsilon_{\q_1+\q_2}-\epsilon_{\q_1}^B-\epsilon_{\q_2}^B)\alpha^{\q_1,\q_2} \\
	E\alpha^{\q_1,\q_2}_\k &= g\gamma^{\q_1,\q_2} + g\sqrt{n_0}\gamma^{\q_1,\q_2}_\k + (\epsilon_{\q_1+\q_2-\k}+\ekb-\epsilon_{\q_1}^B-\epsilon_{\q_2}^B)\alpha^{\q_1,\q_2}_\k \\
	E\alpha^{\q_1,\q_2}_{\k_1,\k_2} &= g (\gamma^{\q_1,\q_2}_{\k_1}+\gamma^{\q_1,\q_2}_{\k_2}) + (\epsilon_{\q_1+\q_2-\k_1-\k_2} + \epsilon_{\k_1}^B + \epsilon_{\k_2}^B - \epsilon_{\q_1}^B - \epsilon_{\q_2}^B) \alpha^{\q_1,\q_2}_{\k_1,\k_2}.
	\end{align}
	\label{eqn:three-body-equations}
\end{subequations}
The impurity operator in Eq.~\eqref{eqn:ansatz-three-body} is normalized according to
\begin{align}
1=\langle \cre{\mathbf{c}}{\0}\ann{\mathbf{c}}{\0}\rangle=&
\abs{\alpha_0}^2 
+ \abs{\gamma_0}^2 
+ \sum_\q f_\q \abs{\gamma^\q}^2 
+ \sum_\q f_\q \abs{\alpha^\q}^2 
+ \sum_\k (1+f_\k) \abs{\alpha_\k}^2 
+ \sum_{\q,\k} f_\q (1+f_\k) \abs{\alpha^\q_\k}^2 
+ \sum_\k (1+f_\k) \abs{\gamma_\k}^2 \nn \\ &
+ \sum_{\q,\k} f_\q (1+f_\k) \abs{\gamma^\q_\k}^2 
+ \frac{1}{2} \sum_{\q_1,\q_2} f_{\q_1} f_{\q_2} \abs{\gamma^{\q_1,\q_2}}^2
+ \frac{1}{2} \sum_{\q_1,\q_2,\k} f_{\q_1} f_{\q_2} (1+f_\k) \abs{\gamma^{\q_1,\q_2}_\k}^2 \nn \\ &
+ \frac{1}{2} \sum_{\k_1,\k_2} (1+f_{\k_1})(1+f_{\k_2}) \abs{\alpha_{\k_1,\k_2}}^2 
+ \frac{1}{2} \sum_{\q,\k_1,\k_2} f_\q (1+f_{\k_1})(1+f_{\k_2}) \abs{\alpha^\q_{\k_1,\k_2}}^2
+ \frac{1}{2} \sum_{\q_1,\q_2} f_{\q_1} f_{\q_2} \abs{\alpha^{\q_1,\q_2}}^2 \nn \\ &
+ \frac{1}{2} \sum_{\q_1,\q_2,\k} f_{\q_1} f_{\q_2} (1+f_\k) \abs{\alpha^{\q_1,\q_2}_\k}^2
+ \frac{1}{4} \sum_{\q_1,\q_2,\k_1,\k_2} f_{\q_1} f_{\q_2} (1+f_{\k_1})(1+f_{\k_2}) \abs{\alpha^{\q_1,\q_2}_{\k_1,\k_2}}^2.
\end{align}

\subsection{Interacting Bose gas}\label{sec:interacting-eqns}

The ansatz used for the interacting Bose gas is,
\begin{equation}
\begin{aligned}
\cre{\mathbf{c}}{\0} ={}& \alpha_0 \cre c\0 + \gamma_0 \cre d\0 + \sum_\q \gamma^\q \cre{d}{\q} \ann\beta \q + \sum_\k \gamma_\k \cre{d}{-\k} \cre \beta \k
+ \sum_\q \alpha^\q \cre{c}{\q} \ann\beta \q + \sum_\k \alpha_\k \cre{c}{-\k} \cre\beta \k + \sum_{\q,\k} \alpha^\q_\k \cre{c}{\q-\k} \cre{\beta}{\k} \ann{\beta}{\q}\\
&+ \frac{1}{2} \sum_{\q_1,\q_2} \alpha^{\q_1,\q_2} \cre{c}{\q_1+\q_2} \ann\beta{\q_1} \ann\beta{\q_2}+ \frac{1}{2} \sum_{\k_1,\k_2} \alpha_{\k_1,\k_2} \cre{c}{-\k_1-\k_2} \cre\beta{\k_1} \cre\beta{\k_2}.
\end{aligned}
\label{eqn:appansatz-bog}
\end{equation}
The resultant system of equations to solve is
\begin{subequations}
	\begin{align}
	E \alpha_0 &= g\sqrt{n_0} \gamma_0 + g \sum_{\q} u_{\q} f_{\q} \gamma^{\q} - g \sum_{\k} v_{\k}(1+f_{\k}) \gamma_{\k} \\
	E \gamma_0 &= g \sqrt{n_0} \alpha_0 + \nu \gamma_0 - g\sum_{\q} v_{\q} f_{\q} \alpha^{\q} + g\sum_{\k} u_{\k}(1+f_{\k}) \alpha_{\k} \\
	E \gamma^\q &= g u_\q \alpha_0 + (\epsilon_{\q,d}+\nu-E_\q)\gamma^\q + g\sqrt{n_0} \alpha^\q + g \sum_{\k} u_{\k} (1\!+\!f_{\k}) \alpha^\q_{\k} - \frac{g}{2} \sum_{\p} v_{\p}f_{\p} (\alpha^{\q,\p}+\alpha^{\p,\q}) \\
	E \gamma_\k &= -g v_\k \alpha_0 + (\epsilon_{\k,d}+\nu+E_\k) \gamma_\k + g\sqrt{n_0} \alpha_\k - g\sum_{\q} v_{\q} f_{\q} \alpha^{\q}_\k + \frac{g}{2} \sum_{\p} u_{\p}(1\!+f_{\p}) (\alpha_{\k,\p}+\alpha_{\p,\k})\\
	E \alpha^\q &= - g v_\q \gamma_0 + g \sqrt{n_0} \gamma^\q + (\eq-E_\q)\alpha^\q \\
	E \alpha_\k &= g u_\k \gamma_0 + g\sqrt{n_0}\gamma_\k + (\ek+E_\k)\alpha_\k \\
	E \alpha^\q_\k &= g u_\k \gamma^\q - g v_\q \gamma_\k + (\epsilon_{\q-\k}+E_\k-E_\q) \alpha^\q_\k\\
	E \alpha^{\q_1,\q_2} &= - g ( v_{\q_2} \gamma^{\q_1} + v_{\q_1} \gamma^{\q_2}) + (\epsilon_{\q_1+\q_2}-E_{\q_1}-E_{\q_2})\alpha^{\q_1,\q_2} \\
	E \alpha_{\k_1,\k_2} &= g ( u_{\k_2} \gamma_{\k_1} + u_{\k_1} \gamma_{\k_2}) + (\epsilon_{\k_1+\k_2}+E_{\k_1}+E_{\k_2})\alpha_{\k_1,\k_2}.
	\end{align}
	\label{eqn:bog-equations}
\end{subequations}
The normalization condition is
\begin{align}
1=&\abs{\alpha_0}^2 
+ \abs{\gamma_0}^2 
+ \sum_\q f_\q \abs{\gamma^\q}^2 
+ \sum_\k (1+f_\k) \abs{\gamma_\k}^2 
+ \sum_\q f_\q \abs{\alpha^\q}^2 
+ \sum_\k (1+f_\k) \abs{\alpha_\k}^2 \nn \\ &
+ \sum_{\q,\k} f_\q(1+f_\k) \abs{\alpha^\q_\k}^2
+ \frac{1}{2}\sum_{\q_,\q_2} f_{\q_1} f_{\q_2} \abs{\alpha^{\q_1,\q_2}}^2 
+ \frac{1}{2}\sum_{\k_1,\k_2} (1+f_{\k_1})(1+f_{\k_2}) \abs{\alpha_{\k_1,\k_2}}^2.
\end{align}

\section{Use of \emph{s}-wave scattering}\label{sec:s-wave}

In this appendix, we discuss how we in practice treat the angles that occur between the multiple momentum vectors in the three-body ansatz, Eq.~\eqref{eqn:ansatz-three-body}.  When three or more momentum vectors are present, then an issue arises with the quadrature of the angular components. To utilize spherical symmetry and thus reduce the number of integrals, one of the vectors is chosen as a reference (i.e., to be aligned with the z-axis) and the other vectors have their direction measured relative to it. However, there are instances where this z-axis vector is removed and the axes must be realigned to another vector. The issue is that for an arbitrary discrete quadrature the transformed coordinates will almost invariably not be the points included in the original quadrature.

To avoid handling explicit angles in the three-body ansatz, we use spherical harmonics, which are discretely indexed and thus avoid the issues which arise from attempting to discretize a continuous sphere. Functions over spherical angles can be decomposed into spherical harmonics, which are a complete set of orthonormal functions on the sphere, as \beq f(\theta,\phi)=\sum_{l=0}^\infty \sum_{m=-l}^l f_{l,m} Y_{l,m}(\theta,\phi), \eeq where $Y_{l,m}$ is a spherical harmonic indexed by $l$ and $m$. In a partial wave expansion, one would sum from $l=0$ to some $l_{max}$. For this work, we take $l_{max}=0$, which means that we only include the isotropic $l=0$ $s$-wave term.  The spherical harmonic $Y_{0,0}(\theta,\phi)=1/\sqrt{4\pi}$ is a constant for all angles.

For our numerical investigations, we expanded the three-body ansatz \eqref{eqn:ansatz-three-body} into spherical harmonics and applied the $s$-wave approximation. This has the net effect of bringing the variational coefficients outside the angular integrals, such that the integration over angle is kept with the operators. To illustrate it with a single term,
\beq
\begin{aligned}
\sum_{\k\ne\q} \alpha^\q_\k \cre{c}{\q-\k} \cre b\k \ann b\q &\rightarrow \int \frac{q^2dq}{2\pi^2}\frac{k^2dk}{2\pi^2}\frac{dx}{2} \alpha^\q_\k \cre{c}{\q-\k} \cre b\k \ann b\q \\
&\rightarrow \int \frac{q^2dq}{2\pi^2}\frac{k^2dk}{2\pi^2} \alpha^\q_\k \int\frac{dx}{2} \cre{c}{\q-\k} \cre b\k \ann b\q,
\end{aligned}
\eeq where in the first line the sum was converted into an integral and spherical symmetry used to simplify the integral (with $x=\cos\theta=\k\cdot\q/kq$), and in the second line the $s$-wave approximation was made to assume that $\alpha^\q_\k$ was independent of angle. The same is done for all other coefficients with angular dependence.  In the variational equations~\eqref{eqn:three-body-equations} only the impurity kinetic energy depends on multiple momentum vectors.  As such, these are the only terms where the integration over angle has an effect.

To consider the angular dependence of the kinetic energy, let us consider four vectors, expressed in spherical coordinates as $\vect{a}=(a,0,0)$, $\vect{b}=(b,\theta_b,0)$, $\vect{c}=(c,\theta_c,\phi_c)$ and $\vect{d}=(d,\theta_d,\phi_d)$. We want to consider the magnitude squared of the sums of these vectors, as they appear in the kinetic energy. By taking dot products, we obtain
\begin{align}
\abs{\vect a+\vect b}^2 &= a^2 + b^2 + 2ab\cos\theta_b, \\
\abs{\vect a+\vect b+\vect c}^2 &= a^2 + b^2 + c^2 + 2a(b\cos\theta_b +c\cos\theta_c) + 2bc(\cos\phi_c\sin\theta_b\sin\theta_c+\cos\theta_b\cos\theta_c),
\end{align}
\begin{align}
\begin{aligned}
\abs{\vect a+\vect b+\vect c+\vect d}^2 ={}& a^2+b^2+c^2+d^2 + 2a(b\cos\theta_b+c\cos\theta_c+d\cos\theta_d)
+ 2bc(\cos\phi_c\sin\theta_b\sin\theta_c+\cos\theta_b\cos\theta_c)\\
& + 2bd(\cos\phi_d\sin\theta_b\sin\theta_d+\cos\theta_b\cos\theta_d)
 + 2cd(\cos(\phi_c-\phi_d)\sin\theta_c\sin\theta_d + \cos\theta_c\cos\theta_d).
\end{aligned}
\end{align}
Integrating out the angular dependence yields
\begin{align}
\int \frac{d\cos\theta_b}{2} \abs{\vect a+\vect b}^2 &= a^2 + b^2,\\
\int \frac{d\cos\theta_b}{2} \frac{d\cos\theta_c}{2}\frac{d\phi_c}{2\pi}\abs{\vect a+\vect b+\vect c}^2 &= a^2 + b^2 + c^2,\\
\int \frac{d\cos\theta_b}{2} \frac{d\cos\theta_c}{2}\frac{d\phi_c}{2\pi} \frac{d\cos\theta_d}{2}\frac{d\phi_d}{2\pi} \abs{\vect a+\vect b+\vect c+\vect d}^2 &= a^2 + b^2 + c^2 + d^2.
\end{align}
These expressions are used in the kinetic energies averaged over angular variables in the three-body equations. Explicitly, in Eq.~\eqref{eqn:three-body-equations} we approximate
\begin{subequations}
\begin{align}
    \epsilon_{\q+\p} &\simeq (q^2+p^2)/2m,\\
    \epsilon_{\q+\p+\k} &\simeq (q^2+p^2+q^2)/2m,\\
    \epsilon_{\q+\p+\k+\l} &\simeq (k^2+p^2+q^2+l^2)/2m.
\end{align}
\label{eqn:averaged-kinetic-energy}
\end{subequations}

\subsection{Demonstration of convergence of the \emph{s}-wave approximation}

Here we characterize the accuracy of the $s$-wave approximation by using the interacting gas ansatz \eqref{eqn:ansatz-bog}, which contained three-body character but was simple enough to allow for angles to be treated numerically exactly. The accuracy was tested using equal impurity and boson masses. The $s$-wave approximation is less accurate for a lighter impurity, because a lighter impurity amplifies the impurity kinetic energy which contains the angular dependence, while it is more accurate for a heavier impurity. The $s$-wave approximation is exact for an infinitely massive impurity, where the associated kinetic energy vanishes.

\begin{figure*}
    \centering
    \includegraphics{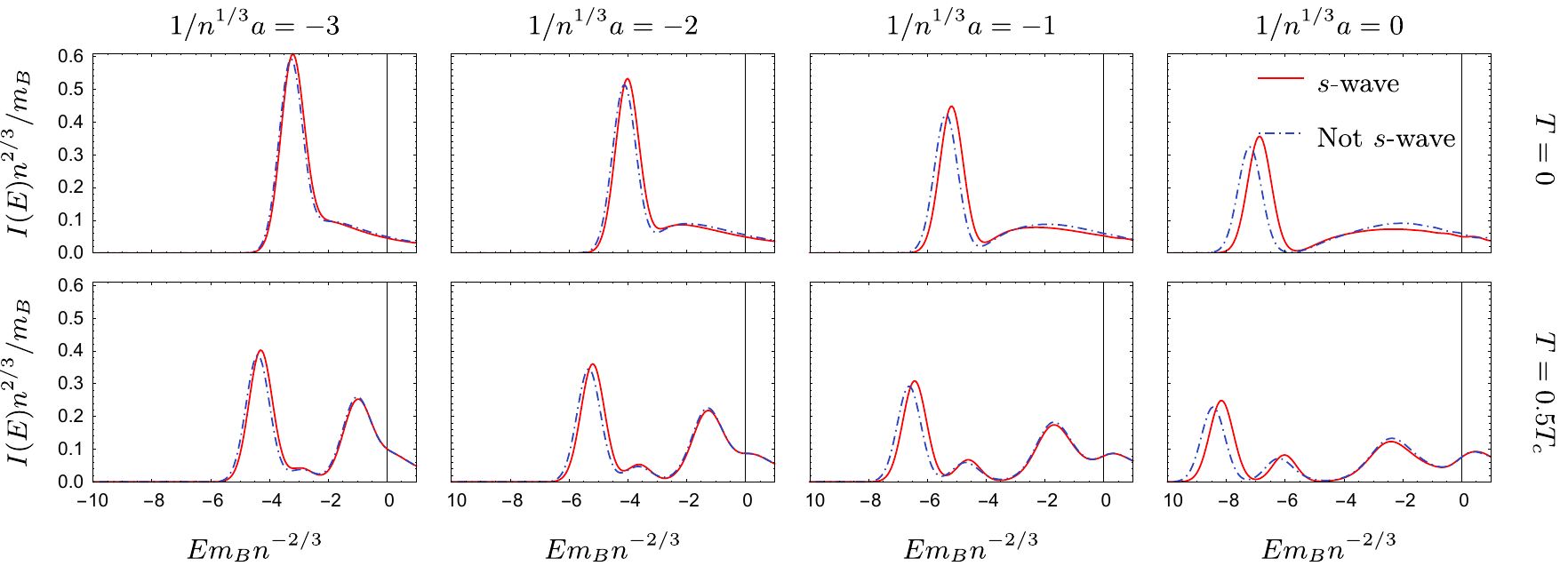}
    \caption{Spectral functions produced using the interacting gas ansatz \eqref{eqn:ansatz-bog}, comparing the $s$-wave approximation (red, solid) with the result of the full angular calculation (blue, dashed) for a variety of different values of $1/a$ and $T$. We take $\ab=0$, $\sigma=0.4 n^{2/3}/m_B$, $R^*=0.02 n^{-1/3}$, and $m=m_B$.}
    \label{fig:s-wave-convergence}
\end{figure*}

We show the accuracy of the $s$-wave approximation in Figure \ref{fig:s-wave-convergence}. We show values of $1/n^{1/3}a$ between $-3$ and 0, at temperatures $T=0$ and $T=0.5T_c$. For $1/n^{1/3}a \lesssim -2$ the $s$-wave approximation gives spectra which are almost indistinguishable from spectra without this approximation.  For stronger interactions, both the energy and amplitude of the attractive polaron and the continuum are somewhat shifted. However, importantly the $s$-wave approximation only changes the quantitative features of the spectrum, while the qualitative features, such as the number of branches, remain unaffected. As such, we expect the $s$-wave approximation to be sufficiently accurate for determining qualitative features of the attractive polaron for $1/a\lesssim 0$. We do not investigate the repulsive polaron in the three-body ansatz because of the difficulty in creating a sufficiently fine integration grid. However, we note that at zero temperature the repulsive polaron is almost fully converged already with one excitation of the medium~\cite{Jorgensen2016}.

\section{Features of the spectral function} \label{sec:spec}

We can gain insight into the behavior of the finite-temperature spectral function by analyzing the effect of omitting different terms from the ansatz.  Figure \ref{fig:drop}(a) shows the spectral function evaluated within the two-body ansatz where some terms including hole excitations of the thermal cloud are omitted, and compares it to the full two-body ansatz.  If we omit either $\alpha^\q$ or $\alpha_\k^\q$ then we see that the doublet splitting of the attractive polaron disappears.  This illustrates how the finite temperature attractive polaron is sensitive to the inclusion of hole excitations of the thermal cloud in the ansatz.

\begin{figure}
    \centering
    \includegraphics{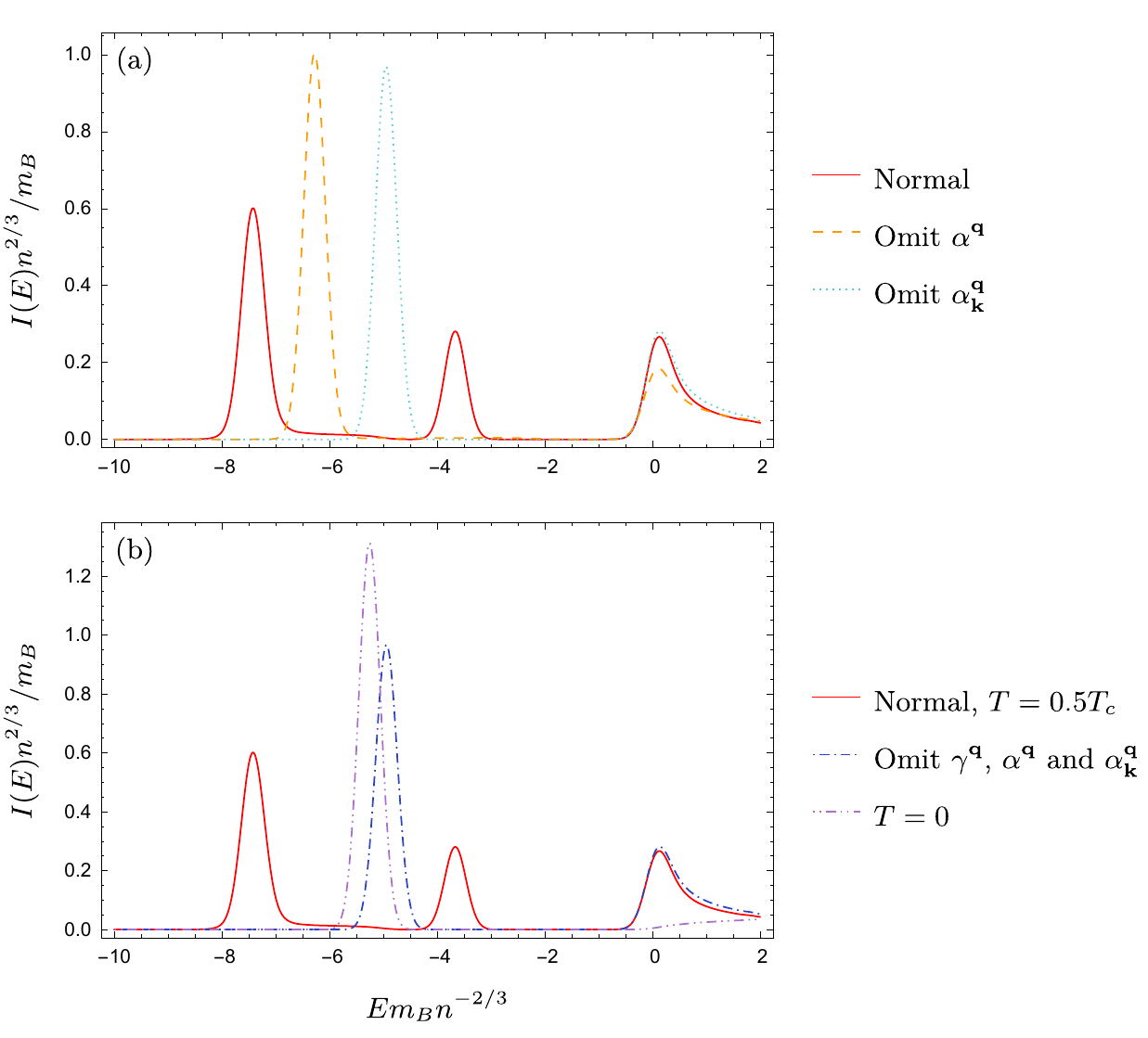}
    \caption{(a) Omitting from the two-body ansatz a term involving scattering of the impurity off thermal bosons makes the splitting disappear, indicating that the splitting arises from scattering off thermal bosons. (b) Omitting all terms involving scattering off thermal bosons does not make the zero-energy peak disappear, indicating that the zero-energy peak is created by some other temperature effect. We take $1/a=0$, $R^*=0.02n^{-1/3}$, $T=0.5T_c$, $m=m_B$ and $\sigma=0.2 n^{2/3}/m_B$.}
    \label{fig:drop}
\end{figure}

Conversely, the peak around zero energy is insensitive to the inclusion or exclusion of hole excitations of the thermal cloud, as shown in Figure \ref{fig:drop}(b).  This zero-energy peak is still sensitive to temperature, and in a BEC changing the temperature changes the thermal cloud.  Since scattering bosons out of the thermal cloud does not effect the zero-energy peak, this implies that the zero-energy peak might be due to bosons being scattered \textit{into} the thermal cloud.  This is supported by inspection of the wavefunction coefficients of the states in this zero-energy peak, as the majority of the weight of each state is in the $\alpha_\k$ term with momentum $\abs{\k}$ close to zero.  This Bose enhancement arises due to the singular nature of the Bose distribution $f_\k$ at low momentum when $T<T_c$.

\begin{figure}
    \centering
    \includegraphics{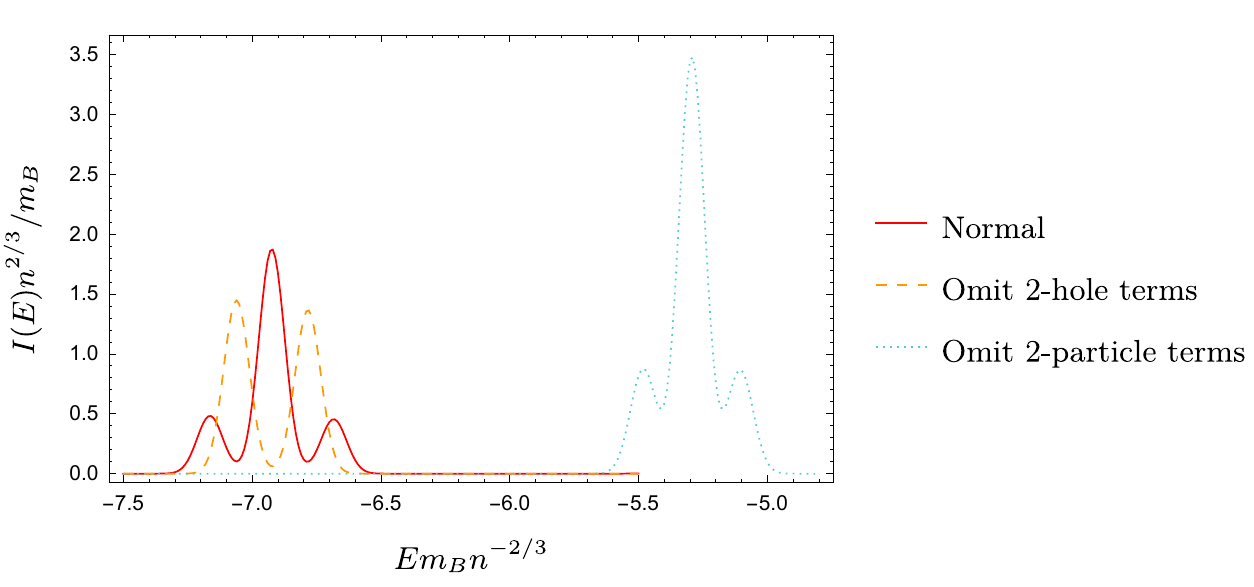}
    \caption{The attractive polaron in the three-body ansatz, where terms involving either two-particle or two-hole excitations have been omitted. We take $1/a=0$, $R^*=0.02n^{-1/3}$, $T=0.01T_c$, $m=m_B$ and $\sigma=0.05n^{2/3}/m_B$.}
    \label{fig:drop3}
\end{figure}

We can extend this analysis to the attractive polaron of the three-body ansatz. We consider what happens to the attractive polaron when we omit either two-particle or two-hole excitations from the ansatz in Figure \ref{fig:drop3}. We see that when we omit the two-hole excitations, we get doublet rather than triplet splitting, in agreement with our general analysis of Sec.~\ref{sec:origin}. When we drop two-particle terms but keep two-hole terms, we retain the triplet splitting. This further supports our finding that the number of splittings of the attractive polaron is dependent on the number of hole excitations.

We have also found that the presence of the continuum is determined by the presence of two-particle terms in the ansatz. Omitting two-particle terms causes the continuum to disappear and the attractive polaron to shift upwards in energy accordingly. Omitting two-hole terms but keeping two-particle terms does not remove the continuum or significantly change the attractive polaron energy.

\begin{figure}
    \centering
    \includegraphics{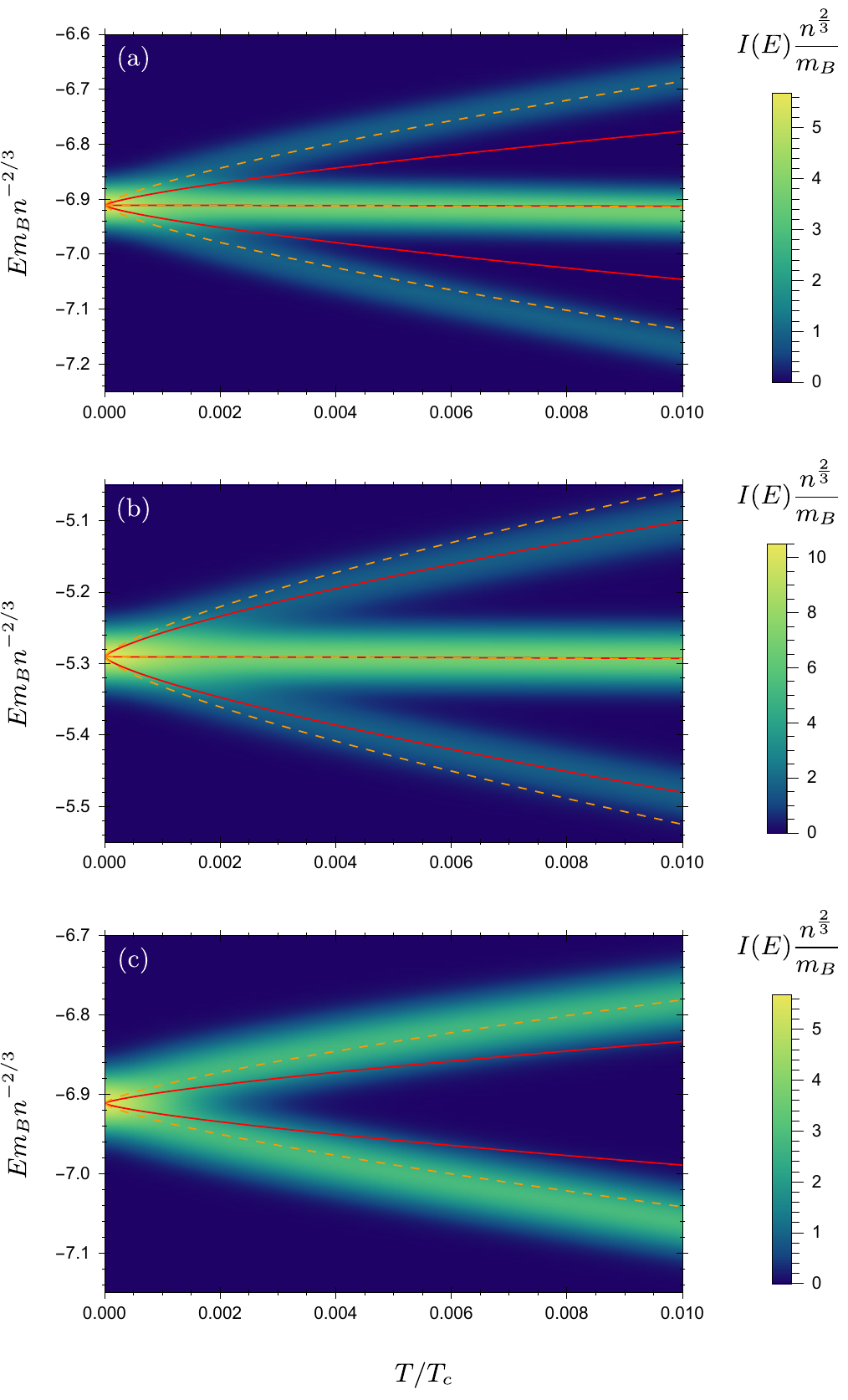}
    \caption{Spectra with respect to temperature for (a) the full three-body ansatz, (b) the three-body ansatz without two-particle terms, and (c) the three-body ansatz without two-hole terms. The solid red lines are the predicted splitting with $Z_{\rm att}$ (\eqref{eqn:pred3Z} for (a) and (b); \eqref{eqn:pred2Z} for (c)). The dashed orange lines are the predicted splitting with $\sqrt{Z_{\rm att}}$ (\eqref{eqn:pred3Sqrt} for (a) and (b); \eqref{eqn:pred2Sqrt} for (c)). These spectra were obtained using $1/a=0$, $R^*=0.02n^{-1/3}$, $m=m_B$ and $\sigma=0.025n^{2/3}/m_B$.}
    \label{fig:Tspec_drop}
\end{figure}

We also investigated how the presence or absence of two-particle and two-hole terms affects the temperature dependence of the splitting compared to the predictions in Eq.~\eqref{eq:DE2body} and \eqref{eq:DE3body}. Direct expansion of the Green's function as we calculated for one-particle terms gives
\begin{equation}
    E\simeq E_{\rm att}\left[1\pm Z_{\rm att} (n_{\rm ex}/n_0)^{1/2}\right]
    \label{eqn:pred2Z}
\end{equation}
for one hole excitation and
\begin{align}
    E\simeq E_{\rm att}\left[1\pm \sqrt{3}Z_{\rm att} (n_{\rm ex}/n_0)^{1/2}\right]
    \label{eqn:pred3Z}
\end{align}
for two hole excitations (with the third solution being $E\simeq E_{\rm att}$). However, if two-particle terms are included, then we have found that using the square root of the residue gives a better fit. For one hole excitation, this is
\begin{equation}
    E\simeq E_{\rm att}\left[1\pm \sqrt{Z_{\rm att}} (n_{\rm ex}/n_0)^{1/2}\right],
    \label{eqn:pred2Sqrt}
\end{equation}
while for two hole excitations this is
\begin{align}
    E\simeq E_{\rm att}\left[1\pm \sqrt{3 Z_{\rm att}} (n_{\rm ex}/n_0)^{1/2}\right].
    \label{eqn:pred3Sqrt}
\end{align}
We compare these predictions against the three-body ansatz in Figure \ref{fig:Tspec_drop}. For the full three-body ansatz, Fig. \ref{fig:Tspec_drop}(a), we find that the square root of the residue \eqref{eqn:pred3Sqrt}, rather than the whole residue, fits very well. However, when we drop two-particle terms in Fig. \ref{fig:Tspec_drop}(b), we get good agreement with the whole residue \eqref{eqn:pred3Z} rather than the square root.

Furthermore, when we drop two-hole terms (but keep two-particle terms) in Fig. \ref{fig:Tspec_drop}(c), we get excellent agreement with the square root of the residue \eqref{eqn:pred2Sqrt}. This is in contrast to the two-body ansatz, which has neither two-hole nor two-particle terms, which fits well with the whole residue \eqref{eqn:pred2Z} (see Fig. \ref{fig:two-body-temperature}).

Therefore, we can infer that whether the splitting scales as the whole residue $Z_{\rm att}$ or the square root of the residue depends on how many particle excitations are included in the impurity operator. A single particle excitation gives $Z_{\rm att}$, while two particle excitations gives $\sqrt{Z_{\rm att}}$. 

\end{widetext}

\bibliography{bosepolaron}

\end{document}